\documentclass{emulateapj}

\usepackage{natbib}
\usepackage{amsmath}

\newcommand{\msun}{M$_{\sun}$}

\newcommand{\ldl}{$\lambda/{\Delta}{\lambda}$}
\newcommand{\teff}{T$_{eff}$}
\newcommand{\logg}{$\log{g}$}

\newcommand{\vsini}{$v\sin{i}$}

\newcommand{\lhalbol}{$\log_{10}{L_{H\alpha}/L_{bol}}$}

\newcommand{\kms}{km~s$^{-1}$}

\newcommand{\masyr}{mas~yr$^{-1}$}
\newcommand{\name}{WISE~J072003.20$-$084651.2}
\newcommand{\namesh}{WISE~J0720$-$0846}

\slugcomment{Submitted to AJ 13 Aug 2014; Accepted for publication 15 October 2014}

\begin{document}

\title{WISE~J072003.20$-$084651.2: An Old and Active M9.5 + T5 Spectral Binary 6 pc from the Sun\footnote{Some of the data presented herein were obtained at the W.M. Keck Observatory, which is operated as a scientific partnership among the California Institute of Technology, the University of California and the National Aeronautics and Space Administration. The Observatory was made possible by the generous financial support of the W.M. Keck Foundation.}}

\author{Adam J. Burgasser\altaffilmark{1,2,3},
Micha\"{e}l Gillon\altaffilmark{4},
Carl Melis\altaffilmark{1}, 
Brendan P. Bowler\altaffilmark{5,6,7}, 
Eric L.\ Michelsen\altaffilmark{1},
Daniella Bardalez Gagliuffi\altaffilmark{1}, 
Christopher R.\ Gelino\altaffilmark{8,9},
E.\ Jehin\altaffilmark{2}, 
L.\ Delrez\altaffilmark{2},
J.\ Manfroid\altaffilmark{2},
\& Cullen H.\ Blake\altaffilmark{10}}

\altaffiltext{1}{Center for Astrophysics and Space Science, University of California San Diego, La Jolla, CA, 92093, USA; aburgasser@ucsd.edu}
\altaffiltext{2}{Visiting Professor at Instituto de Astrof\'{i}sica de Canarias (IAC), La Laguna, Tenerife, Spain}
\altaffiltext{3}{Visiting Astronomer at the Infrared Telescope Facility, which is operated by the University of Hawaii under Cooperative Agreement no. NNX-08AE38A with the National Aeronautics and Space Administration, Science Mission Directorate, Planetary Astronomy Program.}
\altaffiltext{4}{Institute of Astrophysics and G{\'{e}}ophysique, Universit{\'{e}} of Li{\`{e}}ge, all{\'{e}}e du 6 Ao\^{u}t, 17, B-4000 Li{\`{e}}ge, Belgium}
\altaffiltext{5}{California Institute of Technology, Division of Geological and Planetary Sciences, 1200 East California Boulevard, Pasadena, CA 91101, USA}
\altaffiltext{6}{Caltech Joint Center for Planetary Astronomy Fellow}
\altaffiltext{7}{Visiting Astronomer, Kitt Peak National Observatory, National Optical Astronomy Observatory, which is operated by the Association of Universities for Research in Astronomy (AURA) under cooperative agreement with the National Science Foundation}
\altaffiltext{8}{NASA Exoplanet Science Institute, Mail Code 100-22, California Institute of Technology, 770 South Wilson Avenue, Pasadena, CA 91125, USA}
\altaffiltext{9}{Infrared Processing and Analysis Center, MC 100-22, California Institute of Technology, Pasadena, CA 91125, USA}
\altaffiltext{10}{Department of Physics and Astronomy, University of Pennsylvania, Philadelphia, PA 19104, USA}

\begin{abstract}
We report observations of the recently discovered, nearby late-M dwarf WISE~J072003.20$-$084651.2.  New astrometric measurements obtained with the TRAPPIST telescope improve the distance measurement to 6.0$\pm$1.0~pc and confirm the low tangential velocity (3.5$\pm$0.6~{\kms}) reported by Scholz.
Low-resolution optical spectroscopy indicates a spectral type of M9.5
and prominent H$\alpha$ emission ($\langle${\lhalbol}$\rangle$ = $-$4.68$\pm$0.06), but no
evidence of subsolar metallicity or Li~I absorption.
Near-infrared spectroscopy reveals subtle peculiarities that can be explained by the presence of a T5 binary companion, and
high-resolution laser guide star adaptive optics imaging reveals a faint ($\Delta{H}$ = 4.1) candidate source 0$\farcs$14 (0.8~AU) from the primary.  With high-resolution optical and near-infrared spectroscopy, we measure a stable radial velocity of +83.8$\pm$0.3~{\kms}, 
indicative of old disk kinematics and consistent with the angular separation of the possible
companion. 
We measure a projected rotational velocity of {\vsini} = 8.0$\pm$0.5~{\kms}
and find evidence of low-level variabilty ($\sim$1.5\%) in a 13-day TRAPPIST lightcurve, but cannot robustly constrain the rotational period.
We also observe episodic changes in brightness (1-2\%) and occasional flare bursts (4--8\%) with a 0.8\% duty cycle, and 
order-of-magnitude variations in H$\alpha$ line strength. 
Combined, these observations reveal WISE~J0720$-$0846 to be an old, 
very low-mass binary whose components straddle the hydrogen burning minimum mass,
and whose primary is a relatively rapid rotator and magnetically active.
It is one of only two known binaries among late M dwarfs within 10~pc of the Sun, both of which harbor a mid T-type brown dwarf companion.  We show that while this specific configuration is rare (1.4\% probability), roughly 25\% of binary companions to late-type M dwarfs in the local population are likely low-temperature T or Y brown dwarfs. 
\end{abstract}


\section{Introduction}

Stars and brown dwarfs in the immediate Solar Neighborhood ($d < 10$~pc; \citealt{1995AJ....110.1838R,2008AJ....136.1290R,2006AJ....132.2360H,2007AJ....133..439C}) are ideal targets for detailed investigations of the structural, atmospheric and populative properties of these objects. 
This is particularly true for the very lowest-mass (VLM) and lowest-luminosity dwarfs---the late-M, L, T and Y dwarfs---whose recent discovery has been facilitated by wide-area red and infrared imaging surveys (e.g., 2MASS, DENIS, SDSS, UKIDSS, WISE) and multi-epoch red and infrared astrometry (e.g., LSPM, SuperCOSMOS, AllWISE).
Yet, despite their apparent brightness and high proper motion, up to $\sim$20\% of our nearest ($<$20~pc) VLM neighbors remain ``missing'', particularly toward the Galactic plane \citep{2004AJ....128..463R,2005AJ....130.1680L,2006AJ....132.2360H}, as exemplified by the very recent discovery of the third (Luhman~16AB; \citealt{2013ApJ...767L...1L}) and fourth (WISE J085510.83-071442.5; \citealt{2014ApJ...783..122K,2014ApJ...786L..18L}) closest systems to the Sun, comprised of L, T and Y dwarfs.  Several very low-temperature T and Y dwarfs  \citep{2010MNRAS.408L..56L,2010ApJ...718L..38A,2011A&A...532L...5S,2011ApJ...743...50C,2011ApJS..197...19K,2013A&A...557A..43B} and even M dwarfs (e.g., \citealt{2003ApJ...589L..51T,2004AJ....128..437H,2005A&A...435..363D,2005A&A...439.1127S,2006AJ....132.2360H,2014A&A...561A.113S}) have also been uncovered within 5~pc of the Sun in the past decade.

One of the most recent nearby discoveries is {\name} (hereafter {\namesh}), a candidate M/L dwarf 
identified by \citet{2014A&A...561A.113S} in the WISE survey.
With a parallax distance measurement of only 7.0$\pm$1.9~pc, this source had been ``hiding'' in the Galactic plane, its modest proper motion (123$\pm$2~{\masyr}) preventing it from being picked up in earlier astrometric surveys.
\citet{2014A&A...561A.113S} estimated a photometric classificaiton of M9$\pm$1, which was confirmed in near-infrared spectroscopy reported by 
\citet{2014ApJ...783..122K}.  Given its proximity to the Sun, {\namesh} is an important new system for investigating the M dwarf/L dwarf and star/brown dwarf transitions at high spatial, spectral and temporal resolution. 
 
In this article, we report new observations that confirm {\namesh} to be a nearby (6 pc), late-M dwarf, and show it to have old disk kinematics, magnetic activity (including flares), and rapid rotation. We further identify it as a spectral binary system with a T-type brown dwarf companion,
possibly resolved at a separation of $\sim$0.8~AU and estimated orbital period of 2--4~yr.
In Section~2 we describe our observations of the system, including low- and high-resolution optical and near-infrared spectroscopy, high-resolution imaging, and red optical monitoring. 
In Section~3 we analyze these data, determining the optical and near-infrared classifications, improved astrometry, spatial kinematics, rotation, magnetic emission and limits on the photometric variability of this source.
In Section~4 we identify the binary nature of {\namesh}, both as a spectral binary and a potentially resolved system, and characterize its orbital properties from imaging and multi-epoch radial velocity measurements.  
In Section~5 we review the physical properties of this source, and examine it in the context of multiplicity among late M dwarfs in the immediate vicinity of the Sun.
Results are summarized in Section~6.

\section{Observations}

\subsection{Low-Resolution Red Optical Spectroscopy}

\begin{figure}
\epsscale{1}
\plotone{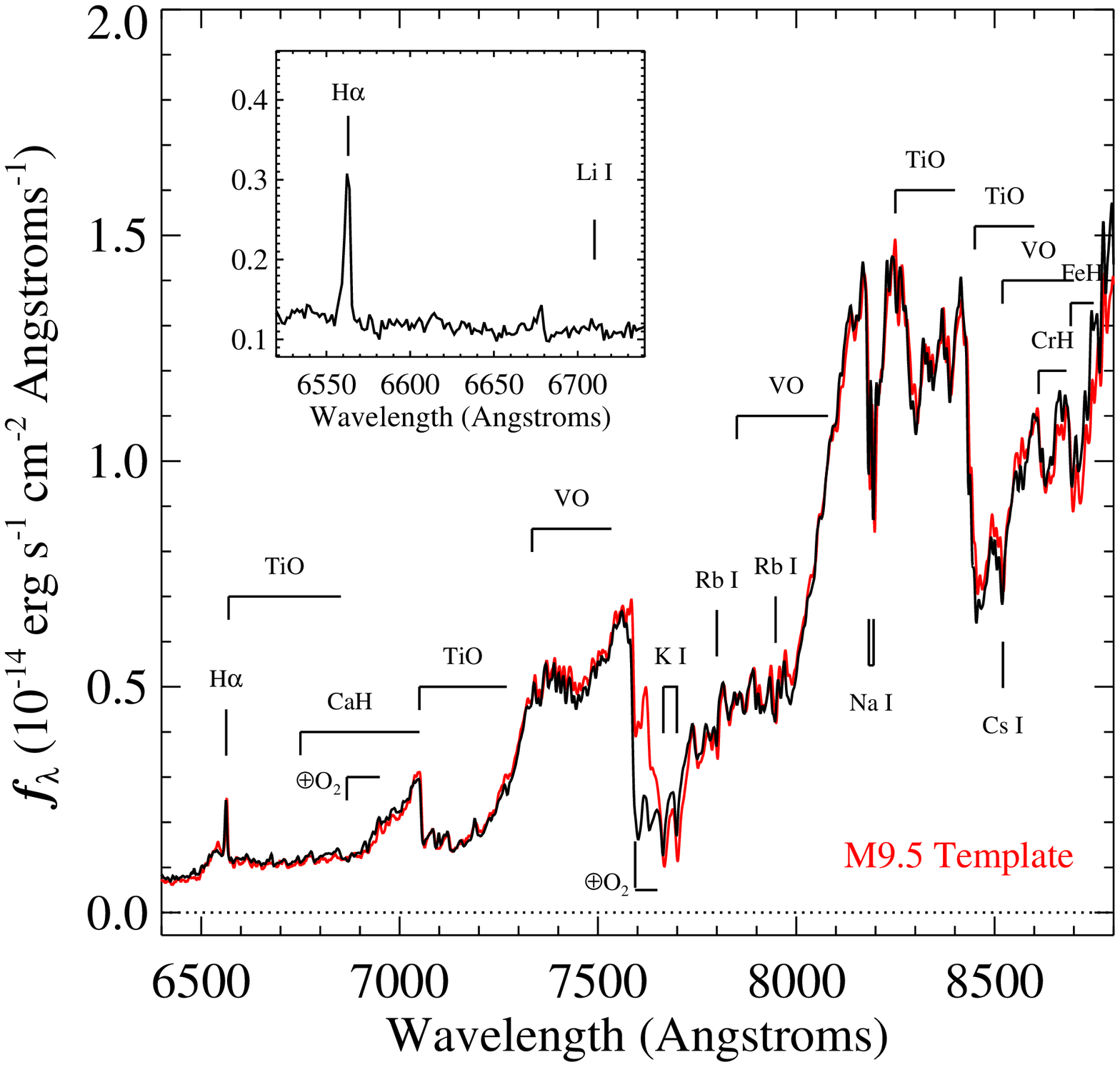}
\caption{Red optical RCSpec spectrum of {\namesh} (black line) scaled to apparent flux units using $R_J$ = 16.80 from SuperCOSMOS \citep{2001MNRAS.326.1279H,2001MNRAS.326.1295H,2001MNRAS.326.1315H}.  The spectrum is compared to an M9.5 spectral template created by merging M9 and L0 SDSS templates from \citet[red line]{2007AJ....133..531B} that is scaled to match {\namesh} in the 7400--7500~{\AA} region. Both spectra are smoothed to a common resolution of {\ldl} = 1200.  Primary atomic and molecular absorption features are labeled, as well as uncorrected telluric ($\oplus$) bands in the {\namesh} spectrum.  The inset box shows a close-up of the 6520--6730~{\AA} region, revealing the presence of H$\alpha$ emission and absence of Li~I  absorption.
\label{fig:optspec}}
\end{figure}

{\namesh} was observed on 2013 December 30 (UT) with the Ritchey-Chretien
Spectrograph (RC Spec) on the Kitt Peak National Observatory's 4m Mayall telescope.  
RC Spec was equipped with the T2KA CCD, and we used the BL420 grating blazed at 7800~{\AA} (first order) and GG-475 order blocking filter with the 1$\farcs$5 $\times$ 98$\arcsec$ slit aligned North to South, yielding spectral data spanning 6300--9000~{\AA} at a resolving power of $\sim$3~{\AA}.  The source was observed near transit at an airmass  of 1.32 in a single integration of 1000~s.  
We also observed the spectrophotometric standard 
HR~3454 \citep{1992PASP..104..533H} at an airmass of 1.15 for flux calibration, as well as flat field and HeNeAr arc lamps. 
Data were reduced using custom Interactive Data Langauge (IDL) routines which corrected images for bad pixels and cosmic rays, performed bias-subtraction, and flat-fielding.  After sky background subtraction and spectral extraction (summing along the spatial axis), we calibrated the wavelength scale to air wavelengths using the arclamp spectrum, and
corrected for throughput losses using the standard star observation. No correction for telluric absorption was attempted.  The reduced spectrum is shown in Figure~\ref{fig:optspec} and discussed further in Section~3.

\subsection{Low-Resolution Near-infrared Spectroscopy}

\begin{figure}
\epsscale{1}
\plotone{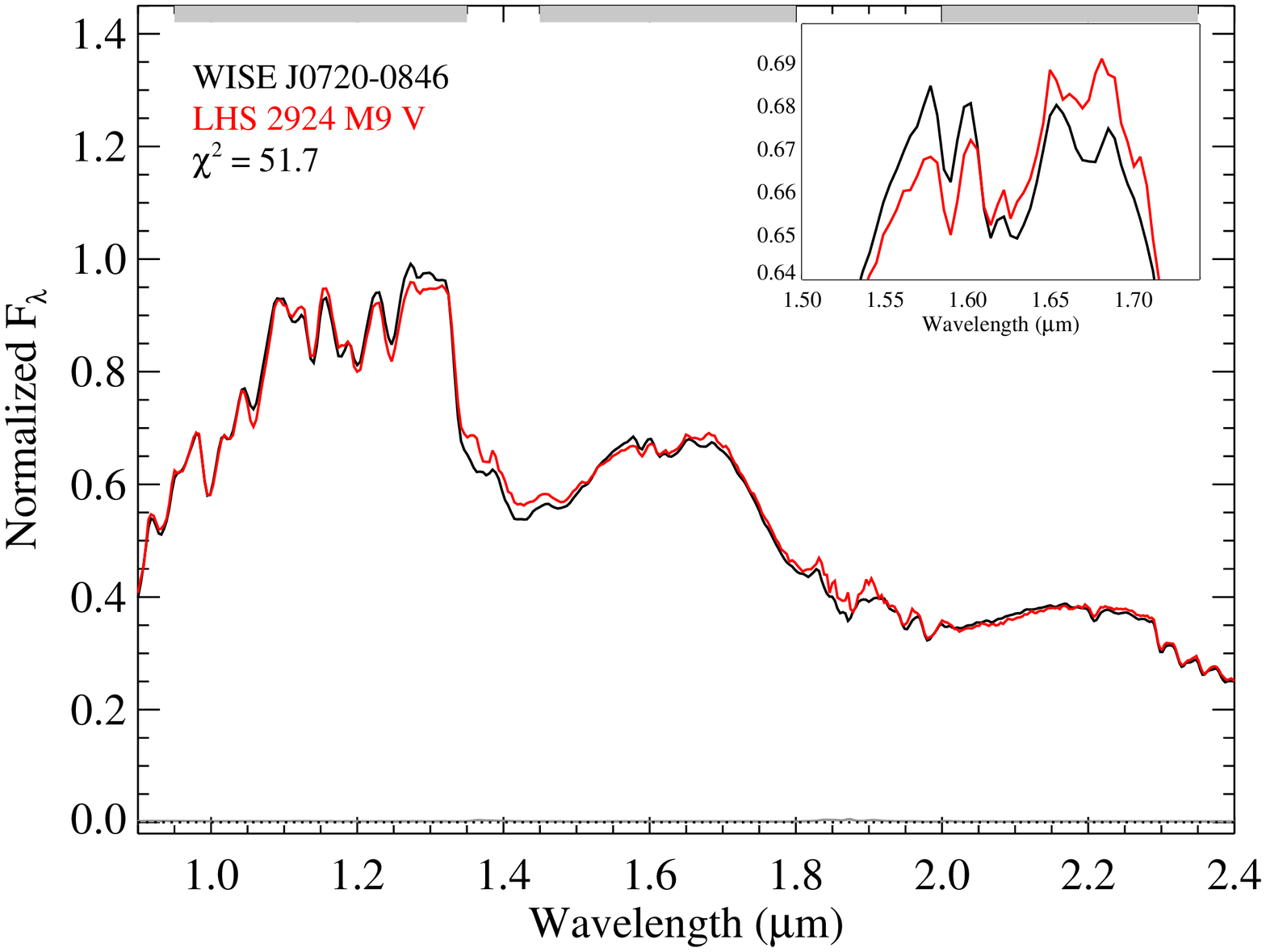}
\plotone{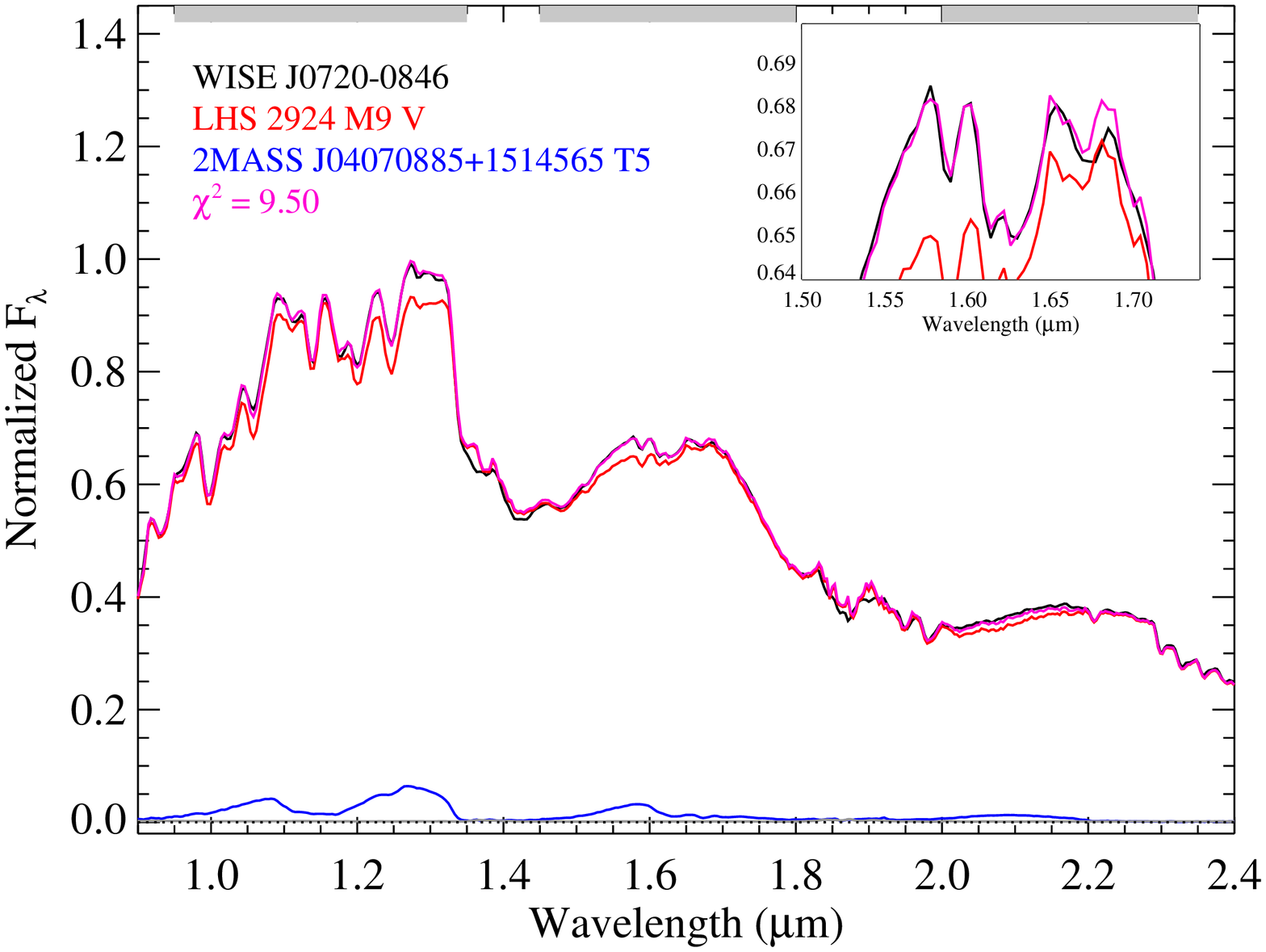} 
\caption{{\em Top:} Normalized SpeX prism spectrum of {\namesh} (black line) compared to the best-fit single source fit from the SpeX Prism Library, the M9 LHS~2924 (red line, data from \citealt{2006AJ....131.1007B}). The inset box focuses on the 1.62~$\micron$ ``dip'' feature in the spectrum of {\namesh}.  Fit regions are indicated by the grey bars at top.
{\em Bottom:} Spectrum of {\namesh} compared to the best-fit binary template, a combination of LHS~2924 (red line) and the T5 2MASS~J04070885+1514565 (blue line; data from \citealt{2004AJ....127.2856B}) relatively normalized according to the $M_K$/spectral type relation of \citet{2007AJ....134.1162L}.   The combined-light spectrum (purple line) coincides well with that of {\namesh}, including replication of the
dip feature and flux excesses at 1.3~$\micron$, 1.6~$\micron$ and 2.1~$\micron$
\label{fig:spex}}
\end{figure}

A low-resolution near-infrared spectrum of {\namesh} was obtained on 2013 December 5 (UT) 
using the SpeX spectrograph mounted on the 3m NASA Infrared
Telescope Facility (IRTF; \citealt{2003PASP..115..362R}).  Conditions were clear with 1$\arcsec$ seeing at $J$-band.  We used the SpeX prism mode with the
0$\farcs$5 slit aligned with the parallactic angle, 
yielding 0.8--2.45 ~$\micron$ spectra
with an average resolution $\lambda/{\Delta}{\lambda}$ $\approx$ 120.
Six exposures of 30~s each were obtained at an airmass of 1.14, followed by observations
of the A0~V star HD~56525 ($V$ = 7.19) at an airmass of 1.20.  
HeNeAr arc lamps and quartz lamp exposures were also obtained for dispersion and pixel response calibration. Data were reduced using the SpeXtool
package version 3.4 \citep{2004PASP..116..362C,2003PASP..115..389V} following
standard procedures for point-source extraction.
These data are similar to those reported by \citet{2014ApJ...783..122K}.
The reduced spectrum is shown in Figure~\ref{fig:spex} and discussed further in Section~4.

\subsection{High Resolution Optical Spectroscopy}

{\namesh} was observed with the Hamilton echelle spectrograph \citep{1987PASP...99.1214V}
on the Lick Observatory Shane 3~m telescope over 7 nights between 21 December 2013 and 26 February 2014 (UT; Table~\ref{tab:hamilton}). Conditions varied from clear to overcast, with seeing between 0$\farcs$8-1$\farcs$5.  We used the 640~$\micron$ slit, 31.5~lines~mm$^{-1}$ grating, and UBK (crown glass) cross-dispersing prisms to obtain 3500--10000~{\AA} spectroscopy over 107 orders,
at a resolution of {\ldl} = 62,000 as measured from the full-width at half-maximum (FWHM) of single arclines in lamp spectra.
Multiple integrations of 40~min were obtained while the source was above an airmass of 2.0.  On each night, halogen lamp flat field frames were obtained for pixel response calibration, and TiAr arclamp spectra were obtained for wavelength calibration.
The M4 radial velocity standard GJ~251 \citep{2002ApJS..141..503N} was observed each night
following WISE\,J0720$-$0846 for
radial velocity calibration.

\begin{deluxetable*}{llccccc}
\tablecaption{Hamilton Spectrograph Observations of {\namesh}\label{tab:hamilton}}
\tabletypesize{\small}
\tablewidth{0pt}
\tablehead{
\colhead{UT Date} &
\colhead{MJD} &
  \colhead{S/N\tablenotemark{a}} & 
 \colhead{V$_{rad}$} &
 \colhead{V$_{rad}$ (H$\alpha$)} &
 \colhead{H$\alpha$ EW} &
 \colhead{\lhalbol} \\
 \colhead{} &
 \colhead{(days)} &
 \colhead{} &
 \colhead{({\kms})} &
 \colhead{({\kms})} &
 \colhead{({\AA})} &
 \colhead{(dex)} \\
}
\startdata
2013 Dec 21 & 56648.02218 & 5 & +82.6$\pm$0.5 & +76$\pm$5 & 2$\pm$1 & $-$5.2$\pm$0.6\\
2013 Dec 22 & 56648.93341 & 15 & +82.5$\pm$0.5 & +82$\pm$5 & 1$\pm$1 & $-$5.9$\pm$1.3 \\
2014 Jan 03 & 56660.96342 & 15 & +82.1$\pm$0.4 & +78$\pm$5 &  6$\pm$1 & $-$4.63$\pm$0.09 \\
2014 Jan 21 & 56678.92950 & 5 & +82.1$\pm$0.5 & +79$\pm$5 & 5$\pm$3 & $-$5.0$\pm$1.0 \\
2014 Feb 23 & 56711.82038 & 25 & +83.0$\pm$0.4 & +79$\pm$5 & 13$\pm$5 & $-$4.3$\pm$0.4 \\
2014 Feb 24 & 56712.80845 & 15 & +82.9$\pm$0.4 & +80$\pm$5 & 5$\pm$2 & $-$4.8$\pm$0.4 \\
2014 Feb 25 & 56713.75047 & 40 & +82.2$\pm$0.3 & +80$\pm$5 & 2$\pm$1 & $-$5.2$\pm$0.6 \\
\enddata
\tablenotetext{a}{Median signal to noise ratio in the 7000--8000~{\AA} region.}
\end{deluxetable*}

Data were reduced in the IRAF environment\footnote{Image Reduction and Analysis Facility \citep{1986SPIE..627..733T}.} following \citet{churchill1995}.
Briefly, data were bias subtracted, flat-fielded, box-car extracted, and wavelength
calibrated using Ti and Ar lines from arclamp exposures \citep{2013AJ....146...97P}.
These data are discussed further in Section~3.

\subsection{High Resolution Near-Infrared Spectroscopy}

High resolution near-infrared $K$-band spectra of {\namesh} were obtained with the NIRSPEC echelle spectrograph on the Keck II telescope \citep{2000SPIE.4008.1048M} on four nights, 
2014 January 19 and 20, March 10 and April 12 (UT; Table~\ref{tab:nirspec}).  
For each night we used the N7 order-sorting filter and 0$\farcs$432-wide slit to obtain 2.00--2.39~$\micron$ spectra over orders 32--38 with {\ldl} = 20,000 ($\Delta{v}$ = 15~{\kms}) and dispersion of 0.315~{\AA}~pixel$^{-1}$. Two dithered exposures of 360--600~s each were obtained, along with observations of the nearby A0~V stars HD~65102 (night~1, $V$ = 6.83) and HD~65158 (nights~2-4, $V$ = 7.16).  Flat field and dark frames were obtained at the start of each night for detector calibration.  

\begin{deluxetable}{llcc}
\tablecaption{Radial and Rotational Velocities from NIRSPEC Observations\label{tab:nirspec}}
\tabletypesize{\small}
\tablewidth{0pt}
\tablehead{
\colhead{UT Date} &
\colhead{MJD} &
\colhead{RV} &
\colhead{\vsini} \\
& &
\colhead{({\kms})} &
\colhead{({\kms})} \\
}
\startdata
2014 Jan 19 & 56676.51  & +84.3$\pm$0.4 & 6.9$\pm$0.7 \\
2014 Jan 20 & 56677.50 &  +83.2$\pm$0.4 & 8.9$\pm$0.7 \\
2014 Mar 10 & 56726.22 &  +84.3$\pm$0.7 & 8.5$\pm$1.2 \\
2014 Apr 12 & 56759.24 &  +83.6$\pm$0.3 & 1.3$\pm$0.7 \\
\enddata
\end{deluxetable}

Data were reduced using a modified version of the REDSPEC package, which took the rectified images produced by that routine and optimally extracted the source spectrum with background subtraction.  We focused exclusively on order~33 (2.29--2.33~$\micron$) which samples CO $\nu$ = 2-0 transitions and telluric CO, H$_2$O and CH$_4$ features. The optimally extracted spectra, which had signals-to-noise exceeding 100 on each night, were forward modeled as described in Section~3.

\subsection{High Angular Resolution Near-Infrared Imaging}

 {\namesh} was observed with the sodium
Laser Guide Star Adaptive Optics system (LGSAO; \citealt{2006PASP..118..310V,2006PASP..118..297W}) and facility Near-InfraRed Camera 2 (NIRC2) on the 10m Keck~II Telescope on 19 January 2014 (UT). Conditions were clear, dry and windy with slightly poor seeing ($\sim$1\arcsec).  
The narrow field-of-view (FOV) camera of NIRC2 was utilized, 
providing an image scale of
$9.963\pm0.011$~mas~pixel$^{-1}$ \citep{2006ApJ...649..389P} over a
10$\farcs$2 $\times$ 10$\farcs$2 area.
We used the MKO\footnote{Mauna Kea Observatories near-infrared filter set \citep{2002PASP..114..169S,2002PASP..114..180T}.} $H$-band filter, and obtained six 60~s integrations using a
three-position dither pattern with variable step size.   
While the LGS provided the wavefront reference source for AO correction, 
tip-tilt aberrations and slow variations were measured by
monitoring the $R=16.8$~mag field star USNO~0812-0137390 located 20$\arcsec$ northeast from {\namesh}.
As this tip-tilt star is faint, and conditions were marginal, we were only able to achieve a Strehl of 1.4\% with these data.

\begin{figure*}
\epsscale{0.8}
\plottwo{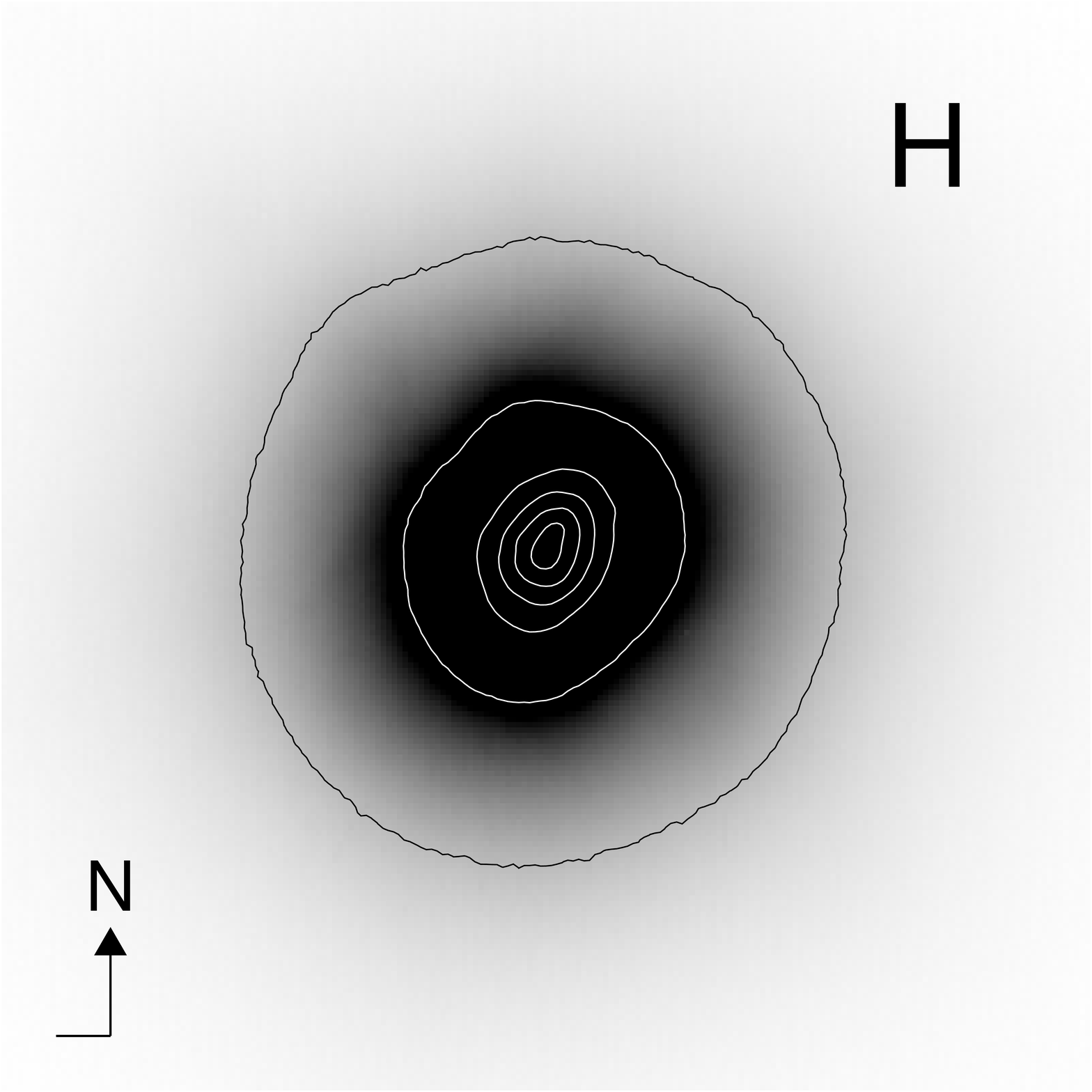}{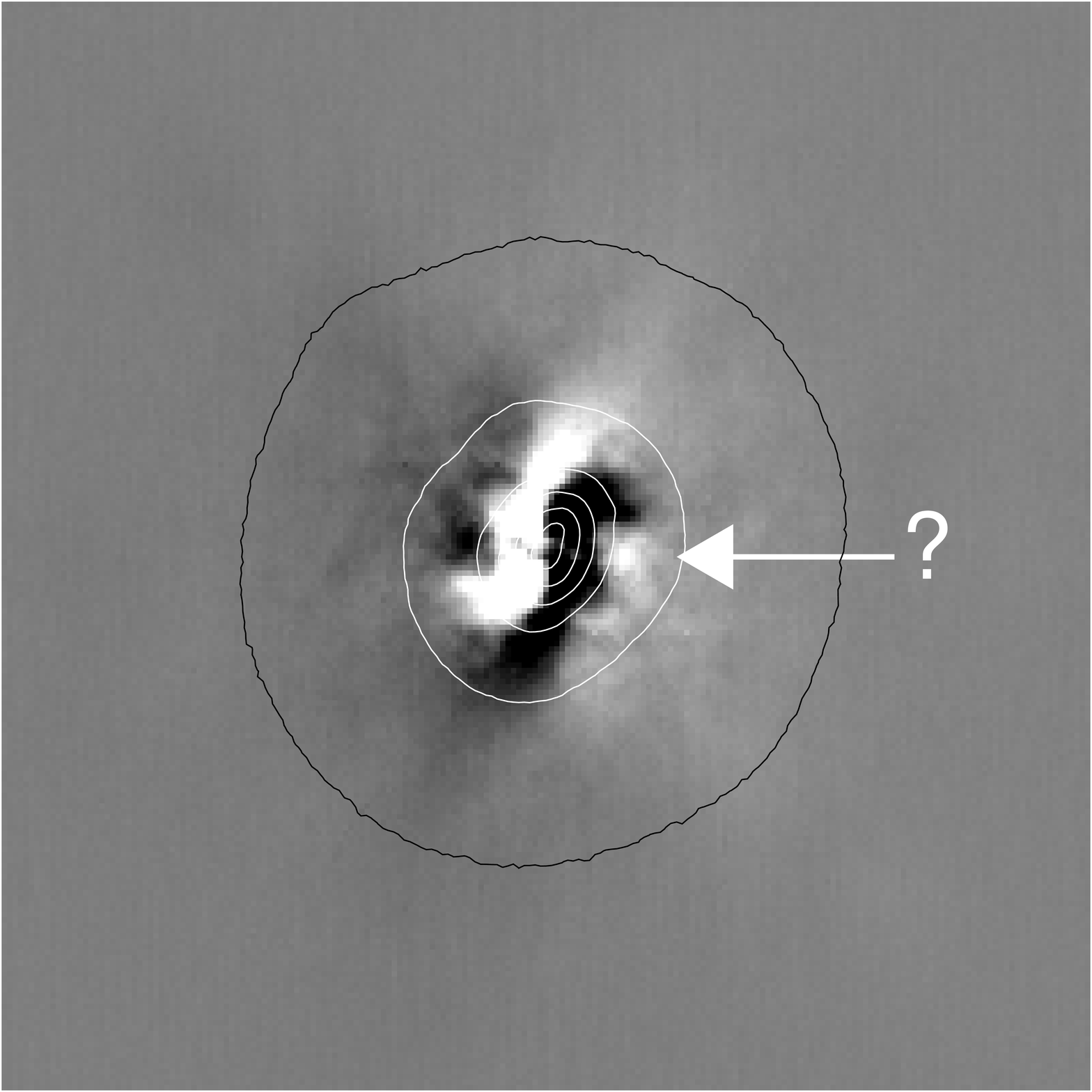} 
\caption{{\em Left}: $H$-band image of {\namesh} taken with NIRC2 LGSAO.  Image is  2$\arcsec$ on a side with North up and East to the left.  White contours indicate 0.1, 0.3, 0.5, 0.7, and 0.9 times the peak flux of the source; black contours indicate the flux level of the predicted companion. 
{\em Right}: Same image but self-subtracted after 180$\degr$ rotation.  A candidate faint source (arrow) is seen west of the primary at a separation of 0$\farcs$14 (1.0~AU projected separation). 
\label{fig:image}}
\end{figure*}

Data were reduced using custom routines to perform flat-fielding, background subtraction, bad-pixel correction, and shifting-and-stacking. The reduced image is shown in Figure~\ref{fig:image}. 
The poor strehl is evident in the broad wings of the point spread function (PSF), which has a azimuthally-averaged FWHM of 0$\farcs$23, and there is a slight elongation perpendicular to the direction of the tip-tilt star. These data are analyzed further in Section~4.

\subsection{Red Optical Photometric Monitoring}

{\namesh} was monitored for 13 non-consecutive days between 30 December 2013 and 16 February 2014 (UT) with the TRAnsiting Planets and PlanetesImals Small Telescope (TRAPPIST; \citealt{2011Msngr.145....2J}), a 0.6~m robotic telescope located at La Silla Observatory in Chile. The telescope is equipped with a thermoelectrically-cooled 2K$\times$2K CCD camera with a 0$\farcs$65 pixel scale and a 22$\arcmin\times$22$\arcmin$ field of view. Light is passed through a broad-band  $I$ + $z$ filter with $>$90\% transmission from 0.75---1.1~$\micron$, the long-wavelength cutoff set by the quantum efficiency of the CCD detector.  Individual exposures of 40~s each were obtained for continuous periods ranging from 4.8 to 8.5 hours, for a total of 73.6~hr on source.  

\begin{figure*}[h]
\center
\epsscale{0.9}
\plotone{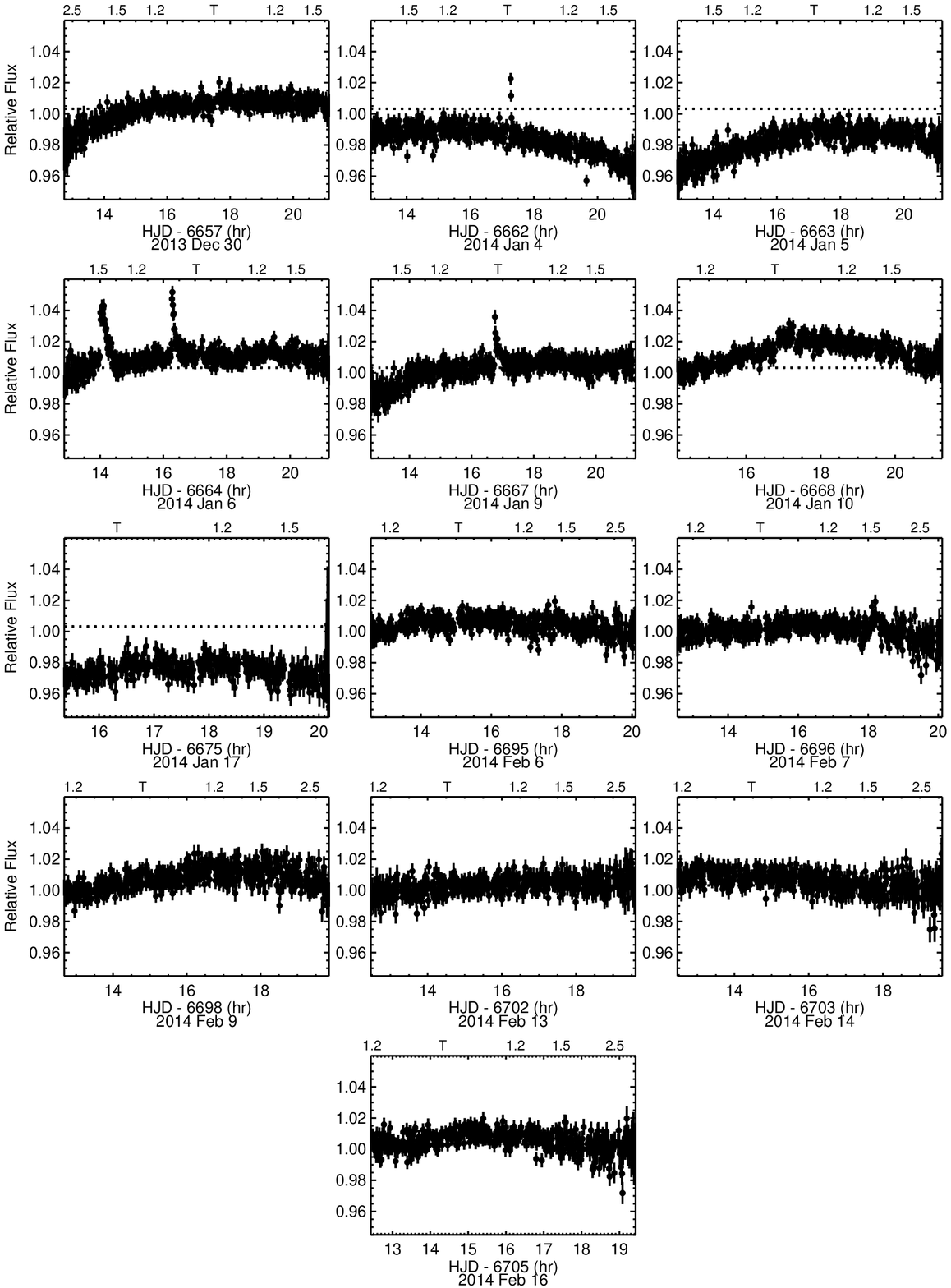}
\caption{TRAPPIST relative light-curves for {\namesh} for each of the 13 nights this source was observed between 30 December 2013 and 16 February 2014 (UT).  Flux values have been normalized to a global mean.  Above each panel the airmasses are indicated, as well as the time of transit ($z$ = 1.067). No correction for airmass-dependent brightness changes has been made in this plot.
}
\label{fig:trappist}
\end{figure*}

\begin{figure*}[h]
\center
\epsscale{0.9}
\plotone{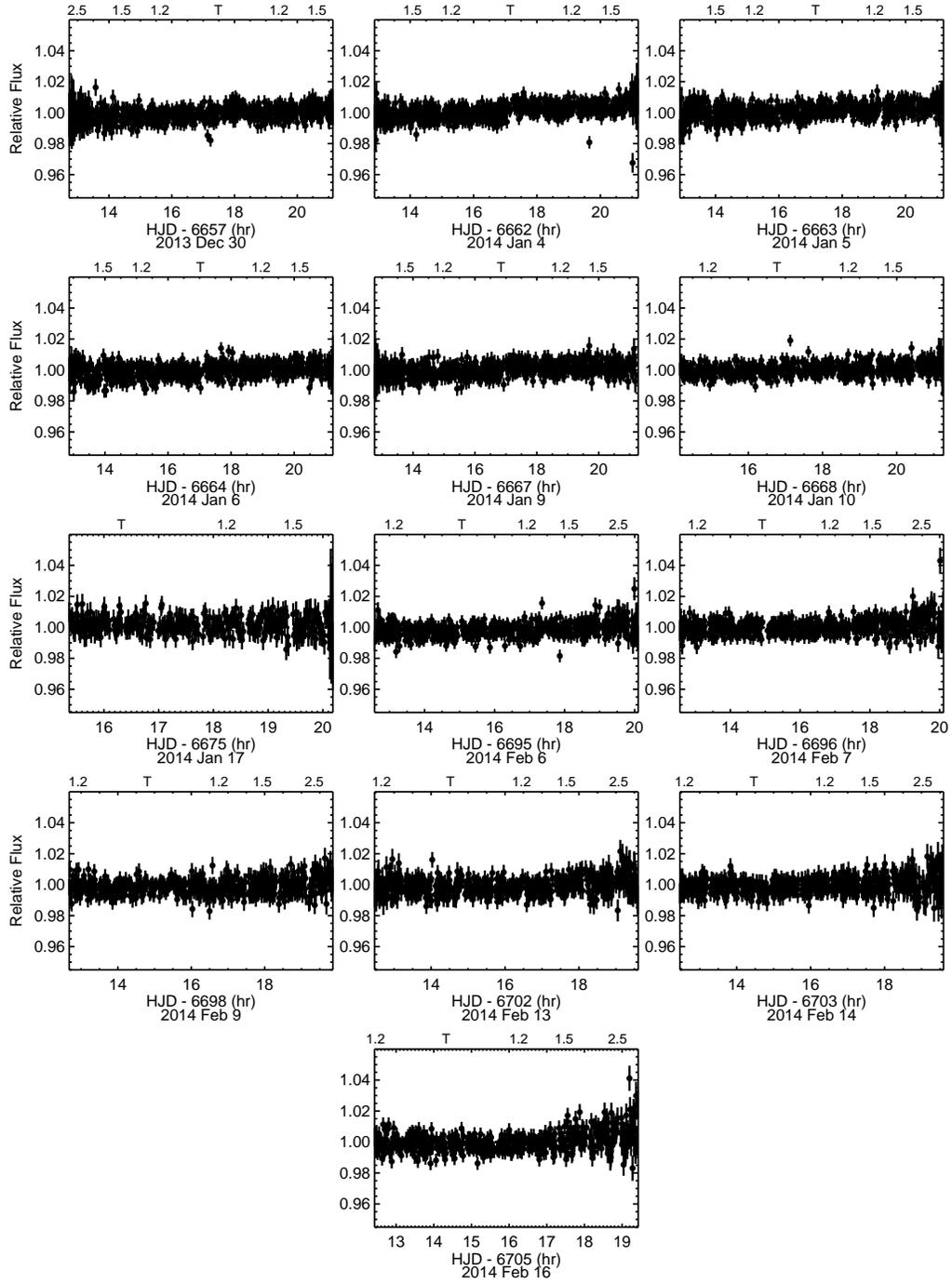}
\caption{Same as Figure~\ref{fig:trappist}, but for the nearby comparison star 2MASS J07200688$-$0846504, 0.06~mag fainter in the TRAPPIST bandpass. 
}
\label{fig:trappist_comp}
\end{figure*}

Data were reduced as described in \citet{2013A&A...555L...5G}.
After a standard pre-reduction (bias, dark, flatfield correction), aperture photometry was performed using IRAF/DAOPHOT2 \citep{1987PASP...99..191S}.  Differential photometry was then determined by comparison to a grid of non-varying background stars, and the overall light curve was normalized to its mean value.  
Relative light curves for {\namesh} and a nearby comparison star 2MASS J07200688$-$0846504 ($J$=12.71$\pm$0.02, $J-K_s$ = 0.39$\pm$0.03) are shown in Figure~\ref{fig:trappist} and~\ref{fig:trappist_comp} as a function of Modified Heliocentric Julian Day\footnote{Julian Day corrected for the Earth-Sun distance, minus 2450000.} (HJD).  These lightcurves are discussed in detail in Section~3.

In addition, nightly astrometry measurements for {\namesh} were determined from the December through February data, as well as a short sequence of 23 frames obtained on 16 September 2014 (UT). To register the astrometric frames, we matched 150 bright, red sources  to the PPMXL catalogue \citep{2010AJ....139.2440R}, using a third-order polynomial (with cross-terms) to map pixels to position.  
Observations on either side of meridian transit had root mean square deviations of order 15~mas; however, a larger shift of order 30-60~mas was found between frames taken before and after meridian. We therefore adopted the mean of pre- and post-meridian measurements as nightly values, and used the cross-meridian shift as an estimate of uncertainty.  These values (equinox J2000) are listed in Table~\ref{tab:astrometry} and discussed further in Section~3.

\begin{deluxetable*}{llccccc}
\tablecaption{TRAPPIST Astrometry\label{tab:astrometry}}
\tabletypesize{\small}
\tablewidth{0pt}
\tablehead{
\colhead{UT Date} &
\colhead{MJD} &
\colhead{$\alpha$} &
\colhead{$\sigma_{\alpha}$} &
\colhead{$\delta$} &
\colhead{$\sigma_{\delta}$} &
\colhead{\# Frames} \\
& &
\colhead{(J2000)} &
\colhead{(mas)} &
\colhead{(J2000)} &
\colhead{(mas)} & \\
}
\startdata
 2014 Jan 4 & 56662.5 & 07$^h$20$^m$03$\fs$2108 & 31 & $-$08$\degr$46$\arcmin$51$\farcs$830 & 28 & 573 \\ 
2014 Jan 5 & 56663.5 & 07$^h$20$^m$03$\fs$2108 & 14 & $-$08$\degr$46$\arcmin$51$\farcs$832 & 34 & 575 \\ 
2014 Jan 6 & 56664.5 & 07$^h$20$^m$03$\fs$2107 & 23 & $-$08$\degr$46$\arcmin$51$\farcs$832 & 36 & 573 \\ 
2014 Jan 9 & 56667.5 & 07$^h$20$^m$03$\fs$2100 & 17 & $-$08$\degr$46$\arcmin$51$\farcs$832 & 33 & 586  \\ 
2014 Jan 10 & 56668.5 & 07$^h$20$^m$03$\fs$2102 & 25 & $-$08$\degr$46$\arcmin$51$\farcs$825 & 36 & 492 \\ 
2014 Jan 17 & 56675.5 & 07$^h$20$^m$03$\fs$2091 & 16 & $-$08$\degr$46$\arcmin$51$\farcs$827 & 43 & 325 \\ 
 2014 Feb 6 & 56695.5 & 07$^h$20$^m$03$\fs$2094 & 11 & $-$08$\degr$46$\arcmin$51$\farcs$821 & 55 & 506 \\ 
 2014 Feb 7 & 56696.5 & 07$^h$20$^m$03$\fs$2060 & 19 & $-$08$\degr$46$\arcmin$51$\farcs$813 & 28 & 506 \\ 
 2014 Feb 9  & 56698.5 & 07$^h$20$^m$03$\fs$2058 & 41 & $-$08$\degr$46$\arcmin$51$\farcs$815 & 28 & 486 \\ 
 2014 Feb 13 & 56702.5 & 07$^h$20$^m$03$\fs$2052 & 29 & $-$08$\degr$46$\arcmin$51$\farcs$809 & 29 & 483 \\ 
 2014 Feb 14  & 56703.5 & 07$^h$20$^m$03$\fs$2063 & 48 & $-$08$\degr$46$\arcmin$51$\farcs$822 & 57 & 480 \\ 
2014 Feb 16 & 56705.5 & 07$^h$20$^m$03$\fs$2045 & 34 & $-$08$\degr$46$\arcmin$51$\farcs$809 & 29 & 472 \\ 
 2014 Sep 17  & 56908.9 & 07$^h$20$^m$03$\fs$2214 & 16 & $-$08$\degr$46$\arcmin$51$\farcs$750 & 16 & 23 \\ 
\enddata
\end{deluxetable*}

\section{System Characterization}

\subsection{Updated Astrometry}

\begin{figure}[h]
\center
\epsscale{1.1}
\plottwo{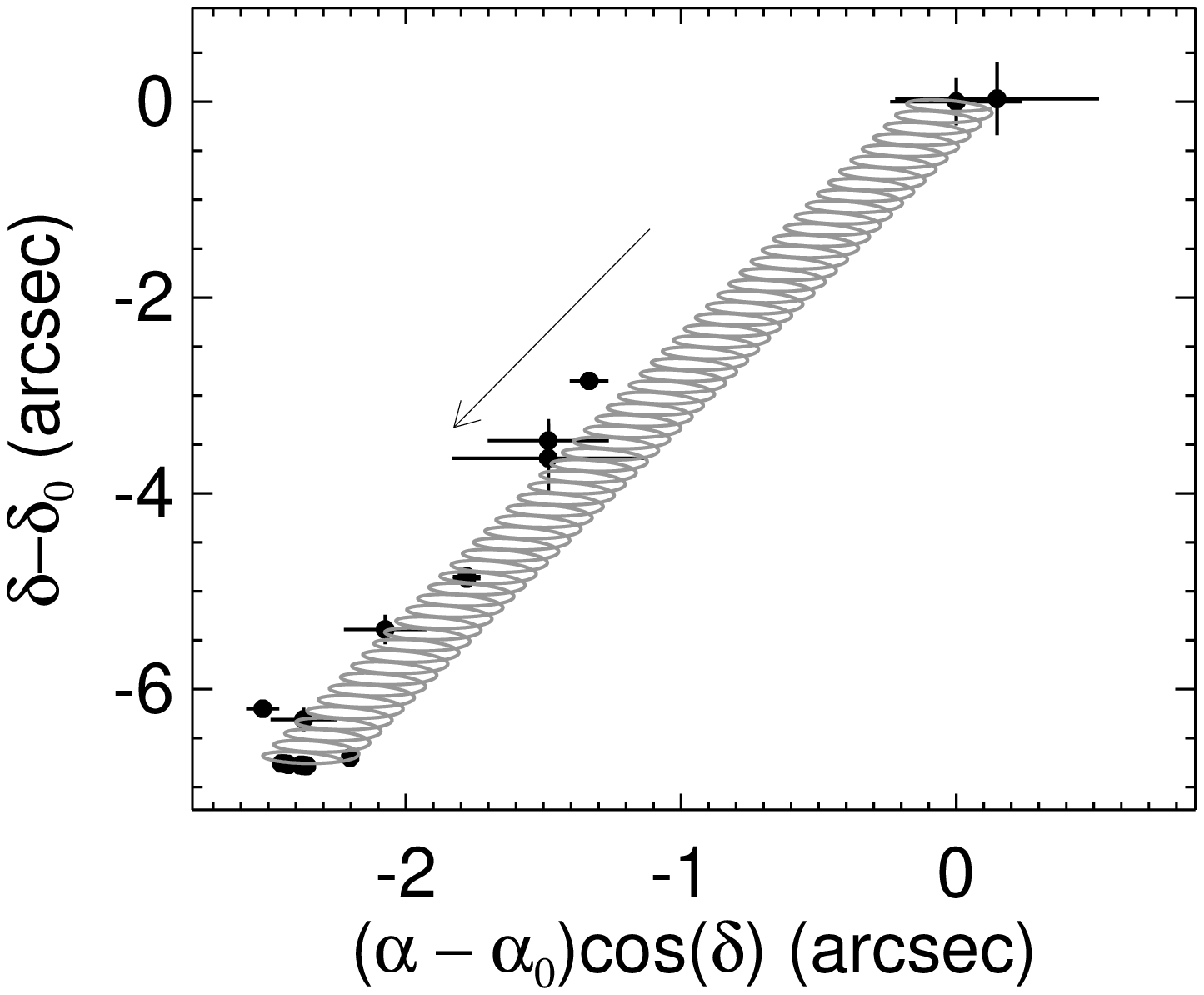}{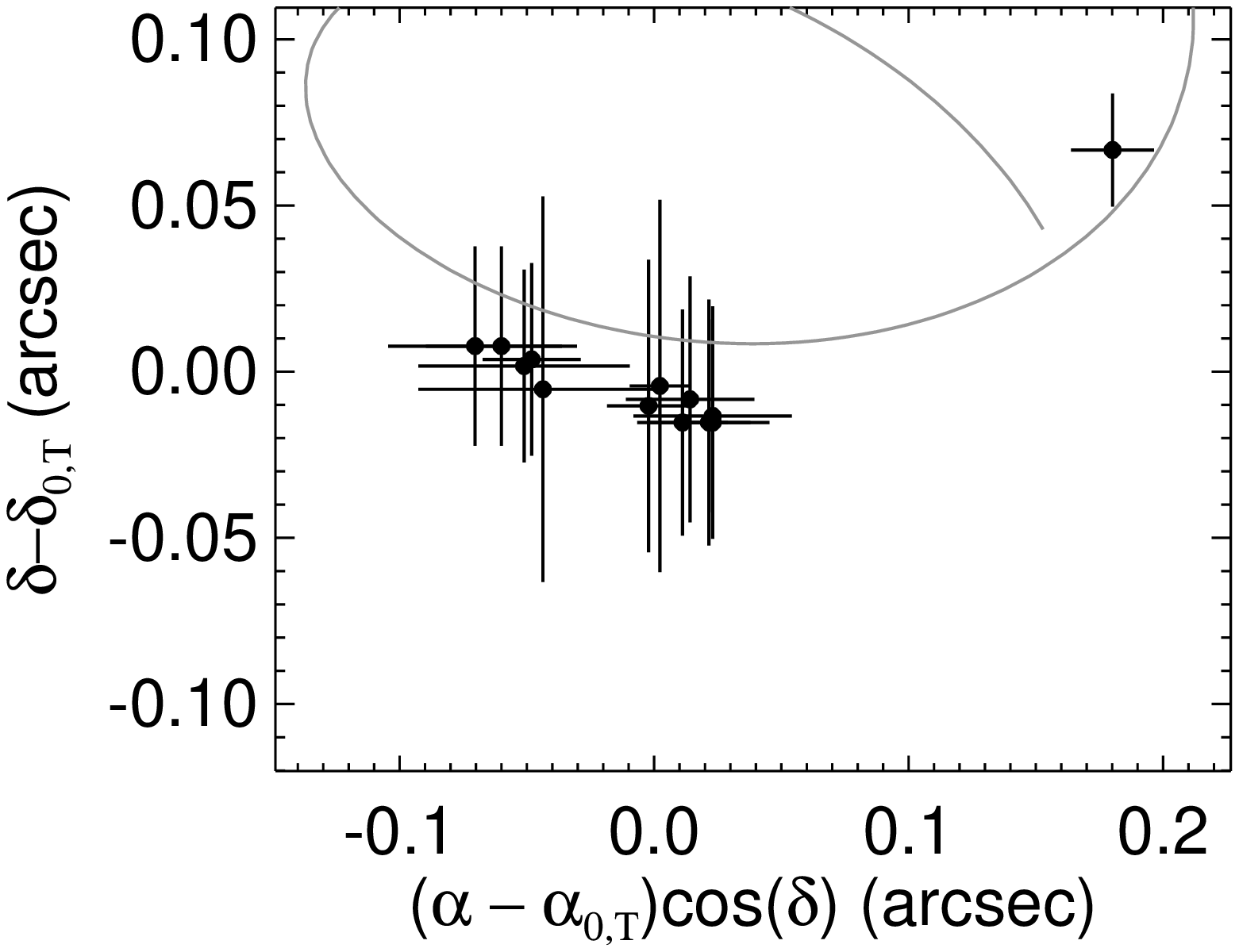}
\caption{(Left): Astrometric solution for {\namesh} (grey line) compared to positional data compiled by \citet{2014A&A...561A.113S} and reported here.  Positions are shown as offsets from a zero point position of 07$^h$20$^m$03$\fs$47 $-$08$\degr$46$\arcmin$36$\farcs$26 at MJD = 35429.0 (1955 Nov 17 UT)  The arrow indicates the general direction of proper motion.  
(Right): Close-up view of TRAPPIST astrometry. 
}
\label{fig:astrometry1}
\end{figure}

\begin{figure}[h]
\center
\epsscale{1.}
\plotone{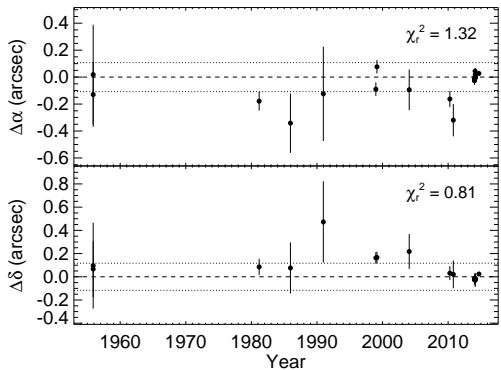}
\caption{Residuals between position measurements and derived astrometric solution in Right Ascension (top) and declination (bottom).  Standard deviations in the residuals are 108~mas and 117~mas, respectively (dotted lines), and reduced $\chi^2$ values are indicated. 
}
\label{fig:astrometry2}
\end{figure}

With 13 nights of precision TRAPPIST astrometry we attempted to validate and improve the parallax and proper motion measurements reported by \citet{2014A&A...561A.113S}. As our new measurements span less than a year, we combined the survey data reported by Scholz with our TRAPPIST measurements, rejecting only the SuperCosmos Sky Survey H$\alpha$ and Short-R survey measurements \citep{2005MNRAS.362..689P} for which the PSFs were reported to be elliptical. A parallax solution was determined using a Monte Carlo Markov Chain (MCMC) analysis with the Metropolis-Hastings sampling algorithm.  Starting from the parameters determined by Scholz (parallax, proper motion) and the average position of the target, we performed a 10$^5$-step random walk, at each step varying the astrometric parameters using normal distributions. We compared the F-distribution probability distribution function for $\chi^2$ residual values computed before and after each step to a uniform distribution as our acceptance ratio.  Ignoring the first 10\% of the chain, we calculated the mean and standard deviation of the parameters in the remaining steps, weighting each solution by the F-distribution.  Our results, displayed in Figures~\ref{fig:astrometry1} and~\ref{fig:astrometry2} and listed in Table~\ref{tab:properties}, are in agreement with \citet{2014A&A...561A.113S} but with improved uncertainties, particularly in proper motion.  Our derived parallax distance, 6.0$\pm$1.0~pc, is more in line with the spectrophotometric estimates reported by Scholz, and the improvement is driven largely by the more accurate TRAPPIST astrometry.  
We aim to continue monitoring this source for further improvement.

\begin{deluxetable}{lcl}
\tablecaption{Derived Properties of the {\namesh}AB System \label{tab:properties}}
\tabletypesize{\small}
\tablewidth{0pt}
\tablehead{
\colhead{Parameter} &
\colhead{Value} \\
}
\startdata
$\alpha$ (J2000)\tablenotemark{a} & 07$^h$20$^m$02$\fs$19  \\
$\delta$ (J2000)\tablenotemark{a} & $-$08$\degr$46$\arcmin$59$\farcs$53  \\
Optical SpT & M9.5   \\
NIR SpT & M9+T5  \\
$\pi$ (mas) & 166$\pm$28  \\
$d$ (pc) & 6.0$\pm$1.0  \\
$\mu_{\alpha}$ (mas~yr$^{-1}$) & $-$40.3$\pm$0.2  \\
$\mu_{\delta}$ (mas~yr$^{-1}$) & $-$114.8$\pm$0.4 \\
$V_{tan}$ ({\kms}) & 3.5$\pm$0.6 \\
RV ({\kms}) & +83.1$\pm$0.4 \\
$U$ ({\kms}) & $-$47.5$\pm$0.4 \\
$V$ ({\kms}) &  $-$47.6$\pm$0.4 \\
$W$ ({\kms}) &  8.0$\pm$0.5 \\
$v\sin{i}$ ({\kms}) & 8.0$\pm$0.5 \\
$\langle${\lhalbol}$\rangle$ & $-4.68{\pm}0.06$ \\
$\Delta{J}$\tablenotemark{b}  &  3.3$\pm$0.2 \\
$\Delta{H}$\tablenotemark{b}  &  4.1$\pm$0.4 \\
$\Delta{K}$\tablenotemark{b}  &  4.7$\pm$0.4 \\
$\rho$ (mas)  & 139$\pm$14 \\
$\rho$ (AU)  & 0.84$\pm$0.17 \\
\enddata
\tablenotetext{a}{At epoch 2014 Jan 1 (UT).}
\tablenotetext{b}{Based on SpeX analysis, magnitudes are on the MKO system.}
\end{deluxetable}

\subsection{Spectral Classification}

The reduced RC Spec spectrum of {\namesh} is compared to an 
M9.5 spectral template in Figure~\ref{fig:optspec}, produced by merging the M9 and L0 SDSS templates from \citet{2007AJ....133..531B}.  This hybrid template provides the best overall match to the optical spectral shape of {\namesh}. Spectral indices from \citet{1999ApJ...519..802K,2003AJ....125.1598L}; and \citet{2007ApJ...669.1235L} are also consistent with this classification and indicate solar metallicity (Table~\ref{tab:indices}).  We detect no 6708~{\AA} Li~I absorption in these data or the co-added high resolution spectrum, to a 3$\sigma$ equivalent width limit of $EW <$ 0.15~{\AA}; we also find no indication of low surface gravity in the strengths of alkali lines or TiO/VO bands \citep{1999AJ....118.2466M,2009AJ....137.3345C}. Both observations imply a mass $\gtrsim$0.06~{\msun} and age $\gtrsim$100~Myr for this source \citep{1997ApJ...482..442B,2008ApJ...689.1295K}.  H$\alpha$ emission is detected, as described below.

\begin{deluxetable}{lcc}
\tablecaption{Optical Classification Indices and Spectral Types for {\namesh}\label{tab:indices}}
\tabletypesize{\small}
\tablewidth{0pt}
\tablehead{
\colhead{Index} &
\colhead{Value} &
\colhead{SpT} \\
}
\startdata
\hline
\multicolumn{3}{c}{\citet{1995AJ....110.1838R}} \\
\hline
TiO5  &   0.463$\pm$ 0.007 &  \nodata  \\
CaH1  &   0.927$\pm$0.020 &  \nodata  \\
CaH2    & 0.454$\pm$0.006 &  \nodata  \\
CaH3   &  0.736$\pm$0.010 &  \nodata  \\
\hline
\multicolumn{3}{c}{\citet{1999ApJ...519..802K}} \\
\hline
CrH-a & 1.091$\pm$0.009 & M9/L0 \\
Rb-b/TiO-b & 0.635$\pm$0.009 & M9/L0 \\
Cs-a/VO-b  &   0.768$\pm$0.008 & M9/L0 \\
\hline
\multicolumn{3}{c}{\citet{2003AJ....125.1598L}} \\
\hline
VO1 &    0.737$\pm$0.006 & M9.7 \\
VO2  &  0.364$\pm$0.003 & M8.6 \\ 
TiO6   &   1.858$\pm$0.017 & \nodata  \\
TiO7   &  0.521$\pm$0.004 & M8.0 \\
Color-M  &    9.93$\pm$0.07 & M9.1 \\
\hline
\multicolumn{3}{c}{\citet{2007ApJ...669.1235L}} \\
\hline
$\zeta$ & 1.034$\pm$0.018 & dM \\
\enddata
\end{deluxetable}

For our near-infrared data, we compared the SpeX spectrum to 727 equivalent spectra of optically-classified M and L dwarfs drawn from the SpeX Prism Library (SPL; \citealt{2014arXiv1406.4887B}).  Following the fitting methodology described in \citet{2010ApJ...710.1142B}, we found a best match to the M9 dwarf LHS~2924 (\citealt{1983ApJ...274..245P}; Figure~\ref{fig:spex}) and a F-distribution weighted mean to all templates of M9.0$\pm$0.7.  Subtle variances between the spectrum of {\namesh} and LHS~2924 are discussed in detail below. These results confirm the near-infrared classification derived by \citet{2014ApJ...783..122K} and establish congruence between the optical and near-infrared spectral morphologies.

\subsection{Kinematics}

\begin{figure*}
\epsscale{0.8}
\plotone{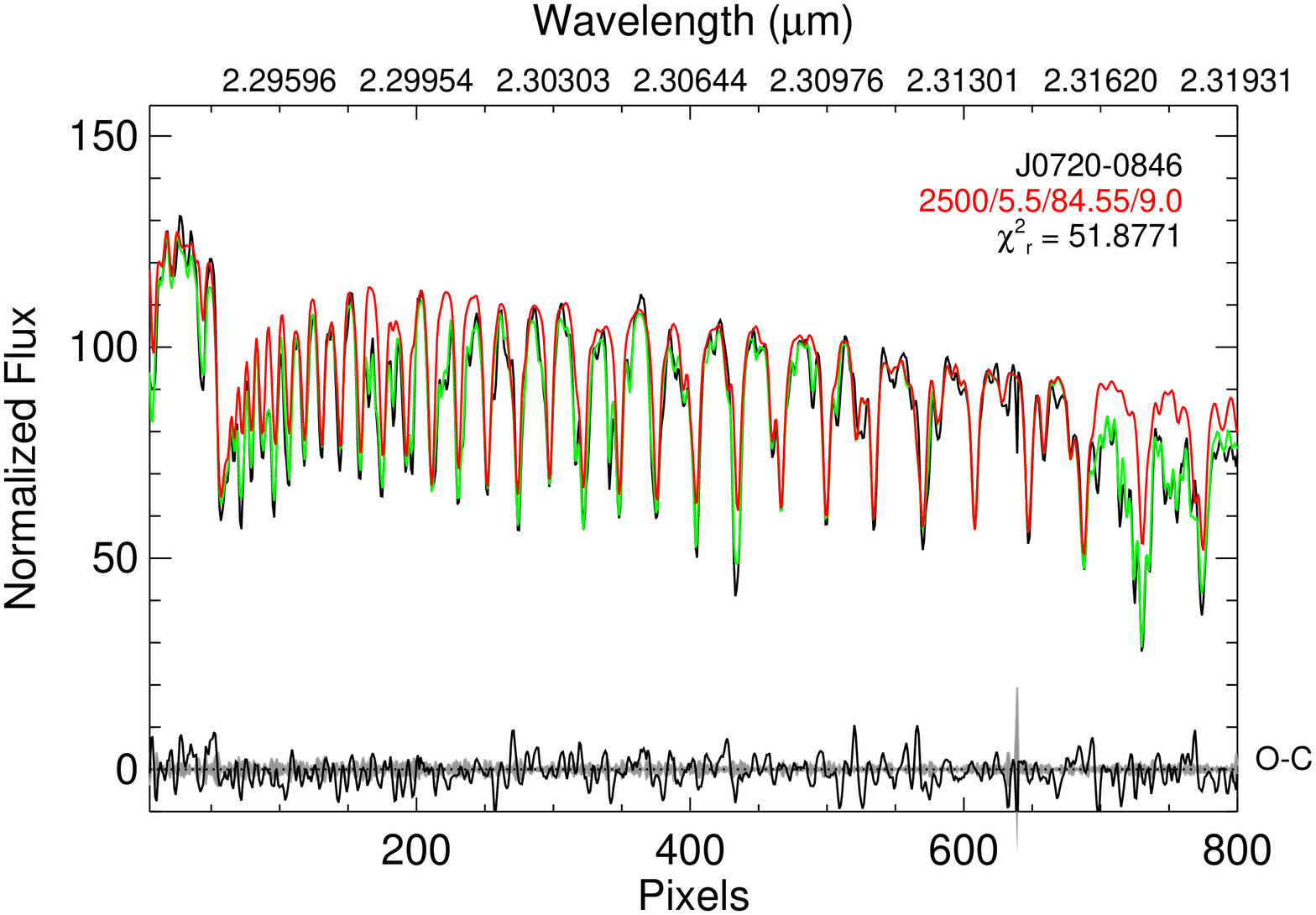} 
\caption{High-resolution ({\ldl} = 20,000) $K$-band spectrum of WISE~J0720$-$0846AB (black line) from 2014 March 10 (UT) compared to a best-fit model combining dwarf spectrum (red line) and telluric absorption (green line) components.  The difference between data and model (O-C)  is shown in black at the bottom of the plot and is dominated by fringing residuals; uncertainty spectrum is indicated in grey.
The forest of CO bandheads allow us to precisely pin down the radial and rotational velocity of the primary of this system.
\label{fig:nirspec}}
\end{figure*}

The small proper motion of {\namesh}, 121.7$\pm$0.3~mas~yr$^{-1}$,
translates into a relatively low tangential velocity, 3.5$\pm$0.6~{\kms}.
We determined the radial velocity of this source from our high-resolution optical and near-infrared data.
For the optical data, we cross-correlated each of the spectra across several orders in the 6600--9000~{\AA} range to the RV standard Gl~251 ($V_r$ = 22.91$\pm$0.10~{\kms}; \citealt{2002ApJS..141..503N}) to determine a relative offset.  The widths of the cross-correlation peak in each order yield velocity uncertainties of 0.5--1.0~{\kms} ($\sim$1/5 of the resolution), which average down to 0.3--0.5~{\kms} after combining all orders.
The NIRSPEC data were forward-modeled using a custom MCMC implementation of the method described in \citet{2010ApJ...723..684B}. The Solar atlas of \citet{1991aass.book.....L} was used to model telluric absorption and the BT-Settl atmosphere models \citep{2011ASPC..448...91A} were used to model the spectrum of {\namesh}.  Spectral models with {\teff} = 2300--2500~K and {\logg} = 5.5 (cgs) provided the best fits.
Figure~\ref{fig:nirspec} shows that this procedure provides an accurate fit to the data, with residuals dominated by uncorrected fringing.  Radial velocity and rotational broadening were among the parameters modeled, and the mean and standard deviation
of their marginalized distributions in the MCMC chain were used as estimates of their measured values and uncertainties. For the radial velocities, uncertainties range over 0.3--0.7~{\kms}. 
Tables~\ref{tab:hamilton} and~\ref{tab:nirspec} list all radial velocity measurements over the 11 epochs observed. 

Both datasets are consistent with a large radial motion
for {\namesh}, with averages of +82.5$\pm$0.4~{\kms} from the Hamilton data and +83.7$\pm$0.4~{\kms} from the NIRSPEC data,\footnote{We attribute the marginal difference (2$\sigma$) between these mean values to a systematic shift between the instruments; see \citet{2005AJ....129.1706F}.} in clear contrast to its small tangential motion. The total spatial velocity vector in the Local Standard of Rest\footnote{Assuming a right-handed Cartesian coordinate system centered on the Sun with $U$ pointed radially away from the Galactic center, $V$ in the direction of Galactic rotation, and $W$ toward the Galactic north pole.  We also adopt a Local Standard of Rest motion of ($U$,$V$,$W$)$_{LSR}$ =(11.1, 12.24, 7.25)$\pm$(0.7, 0.5, 0.4)~{\kms} \citep{2010MNRAS.403.1829S}.} is ($U$,$V$,$W$) = ($-$47.5, $-$47.6, 8.0)$\pm$(0.4, 0.4, 0.5)~{\kms}, which falls well outside the 1$\sigma$ distribution
of local late-type M dwarfs, ($\sigma_U$,$\sigma_V$,$\sigma_W$) = (32,20,17)~{\kms} \citep{2002AJ....124.2721R,2009ApJ...705.1416R}.  Its Galactic orbit, computed as described in \citet{2009ApJ...697..148B}, is modestly eccentric, spanning Galactic radii of 4--9~kpc (assuming a Solar radius of 8.5~kpc).
Using the criteria of \citet{1992ApJS...82..351L}, we assign this source to the old disk kinematic population, suggesting a ``mature'' age of 0.5--5~Gyr \citep{1969PASP...81..553E}.

\subsection{Magnetic Activity}

H$\alpha$ emission is detected in both the RC Spec and Hamilton Spectrograph data.  In the low-resolution data, we measure an equivalent width $EW$ = $-$5.2$\pm$0.3~{\AA}.  Using a $\log_{10}{\chi} \equiv \log_{10}{f_{6560}/f_{bol}} = -5.4{\pm}0.1$ from \citet{2004PASP..116.1105W}, we determine {\lhalbol} = $\log_{10}{(\chi{EW})}$ = $-4.68{\pm}0.06$, which is typical for M9 dwarfs in the vicinity of the Sun \citep{2007AJ....133.2258S,2011AJ....141...97W}.

\begin{figure}
\plotone{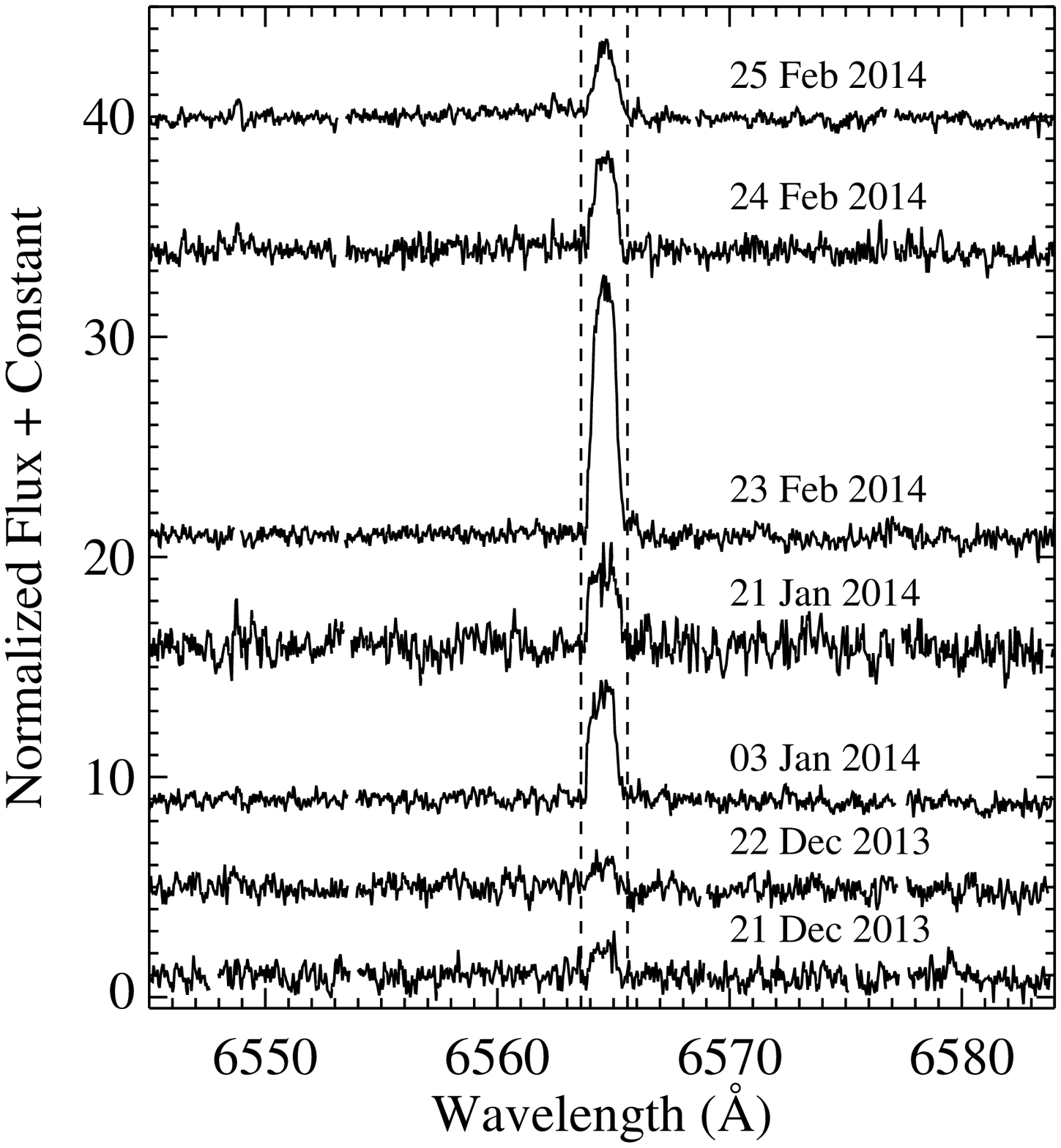}
\caption{
Lick Hamilton spectra of {\namesh} showing the H$\alpha$ emission feature.
All spectra are normalized such that the median continuum around the H$\alpha$
line is unity; the spectra are then offset vertically for clarity. Variability in the 
amplitude of the H$\alpha$ emission is clear. The dashed vertical lines encompass a region
with velocity width of $\approx$90~{\kms}.
Wavelengths in this figure are heliocentric and plotted in
air. Some residual sky lines have been
removed manually from the spectra. 
\label{fig:halpha}}
\end{figure}

Figure~\ref{fig:halpha} displays our Hamilton
spectral data of {\namesh} around the 6563~{\AA} H$\alpha$ line.
Continuum emission is weakly detected at red wavelengths in these data, 
but the H$\alpha$ line in unambiguous, with both broad and highly variable emission.
Equivalent widths (absolute values) range from $\lesssim$1 to 13$\pm$5~{\AA}, corresponding to {\lhalbol} $\lesssim-6$ (inactive) to $-4.3$ ($>$2$\times$ ``normal'' activity).
These lines are also well-resolved, with FWHM of $\sim$40--60~{\kms},
over 5 times the NIRSPEC-derived rotational velocity (see below). 
The H$\alpha$ line centers are nevertheless aligned with absorption line Doppler shifts to within 5~{\kms}, ruling out jets or accretion projected along the line of sight.

\begin{figure}[h]
\center
\epsscale{1.0}
\plotone{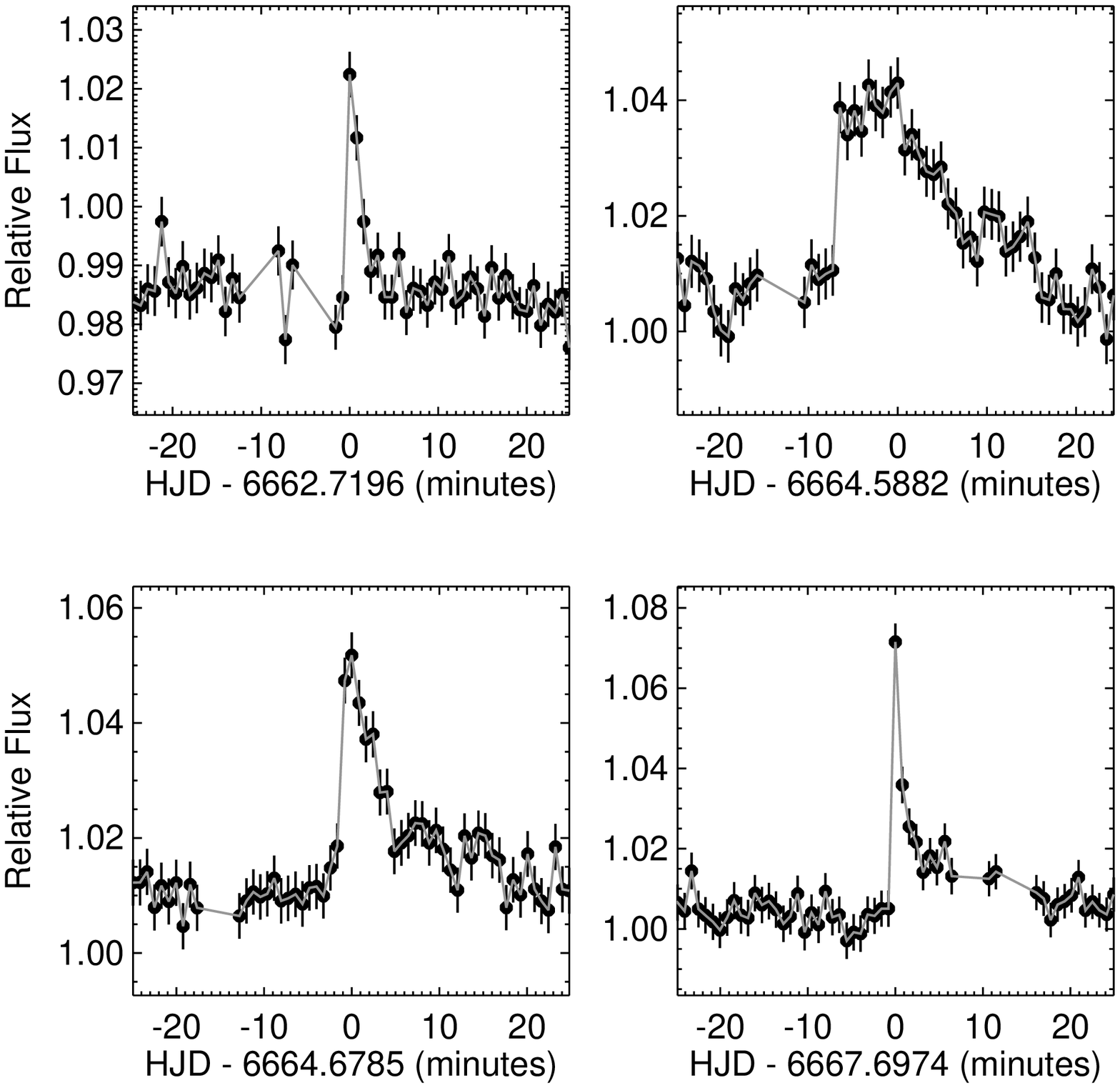}
\caption{Individual broadband flares detected in the TRAPPIST light curve of {\namesh}.
Note the variation in vertical scale in each panel.  Three of the bursts
show a classical burst profile, with an impulsive rise and exponential decay.  
The burst at HJD 6664.59, which occurs 2.2~hr before the burst at HJD 6664.68,
is more extended in time, and may represent sequential flares or an extended, active spot.
}
\label{fig:flare}
\end{figure}

In addition to emission line variability, the TRAPPIST light curve shows several impulsive bursts on HJD 6662, 6664 and 6667, visible in Figure~\ref{fig:trappist} and shown in detail in Figure~\ref{fig:flare}. 
These have relatively low power, with peak fluxes 4--8\% above the local continuum.  Three exhibit
classic flare profiles, with impulsive rises ($<$ 1~min) followed by exponential declines (2--5~min). 
In contrast, the burst on HJD 6664.5859, one of two in a 2~hr period, shows a more complex temporal structure,
with a broad peak that persists for over 5~min followed by a 15~min decline with possible secondary bursts
after 10~min. This may reflect an associated chain of flare events over an extended region, or a massive flare rotating out of view.
These events indicate a 0.8\% duty cycle of flare emission at the level of 2\% above quiescence, somewhat below the typical duty cycles of late M dwarfs (3$\pm$1\%; \citealt{2010AJ....140.1402H}).

Hence, while {\namesh} exhibits numerous signs of activity, including an unusually broad H$\alpha$ line, the strength of its persistent emission and frequency of flaring are at or below the average of equivalently classified dwarfs.  This is consistent with its old kinematic age and slow rotation rate (see below), although it should be noted that age-activity and rotation-activity trends are not well established in the late M and L dwarf regime \citep{2000AJ....120.1085G,2003ApJ...583..451M,2008ApJ...684.1390R,2012ApJ...746...23M}. 

\subsection{Rotation}

In addition to radial motion, our NIRSPEC analysis provides multi-epoch measurements of rotational broadening. We measure a consistent {\vsini} = 8.0$\pm$0.5~{\kms}, which is on the low end of, but consistent with, the rotational velocities of equivalently classified dwarfs \citep{2008ApJ...684.1390R,2010ApJ...723..684B,2012ApJ...746...23M}. Assuming a radius of 0.1~$R_{\sun}$ \citep{2001RvMP...73..719B}, this velocity corresponds to a maximum rotation period of 15$\pm$1~hr, consistent with period measurements for equivalent-mass objects \citep{2011ApJ...727...56I}. Despite being on the low end of the {\vsini} distribution of late-type M dwarfs, this source is still a rapid rotator, a likely explanation for its magnetic activity.

\subsection{Search for Photometric Variability\label{sect:variability}}

To confirm the slow rotation of {\namesh}, we searched for rotational modulation of surface features in the TRAPPIST lightcurve. As shown in Figure~\ref{fig:trappist}, the red optical brightness of this source exhibits considerable night-to-night variation, particularly prior to HJD 6695, the same period we detect flaring bursts. From HJD 6695 onward, {\namesh} is far more stable. Hereafter, we refer to these periods as the ``active'' and ``quiescent'' phases.  For comparison, the relative fluxes of the nearby comparison star (0.06~mag fainter in the TRAPPIST bandpass) are steady throughout the observing period (Figure~\ref{fig:trappist_comp}). 

\begin{figure}[h]
\center
\epsscale{1.1}
\plotone{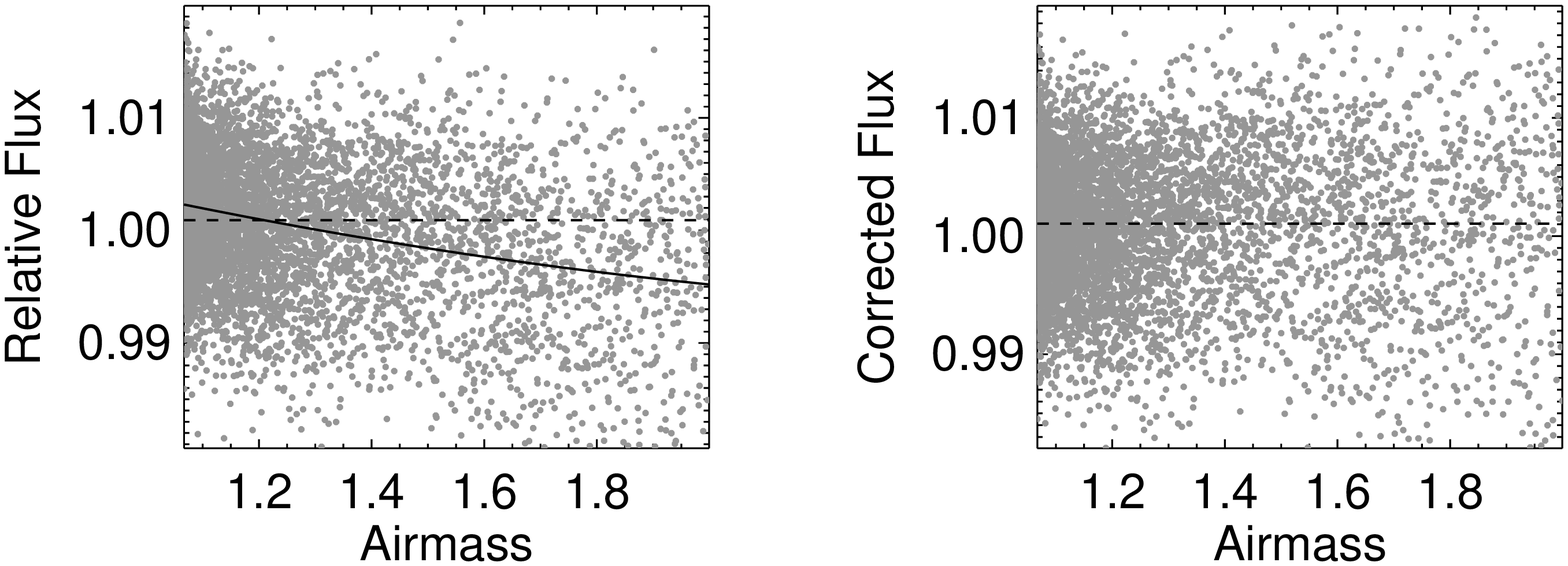}
\caption{TRAPPIST relative fluxes for {\namesh} as a function of airmass, before (left) and after (right) correction of a second-order airmass term (solid line in the left panel). Flares have been removed from these data. The significant trend is likely due to uncorrected color extinction at high airmass resulting from the very different spectral energy distribution of {\namesh} compared to nearby comparison stars.
}
\label{fig:airmass}
\end{figure}

The nightly variations appear to be episodic, and we could find no clear period associated with them.  We therefore normalized each night's light curve to search for persistent low-level variability.
There is additional structure in the normalized lightcurves that arises from differential color extinction at large airmass \citep{2003MNRAS.339..477B}. {\namesh} has a distinct spectral energy distribution in the 0.75-1~$\micron$ TRAPPIST band compared to its neighboring sources, so color-dependent extinction is not corrected in relative photometry.  Figure~\ref{fig:airmass} shows source flux as a function of airmass. A statistically significant trend is present, reaching a deviation of 0.5\% at an airmass of 2.0. In contrast, no significant trend is found in the comparison source.  This airmass effect was corrected in the lightcurve of {\namesh} by dividing by a second order polynomial fit to airmass; we also rejected all measurements at airmasses greater than 2.0 to minimize residual bias. This reduced our total number of samples to 5895.  We found no significant trends with airmass or seeing after the airmass correction was applied.

\begin{figure}[h]
\center
\epsscale{0.8}
\includegraphics[width=0.45\textwidth]{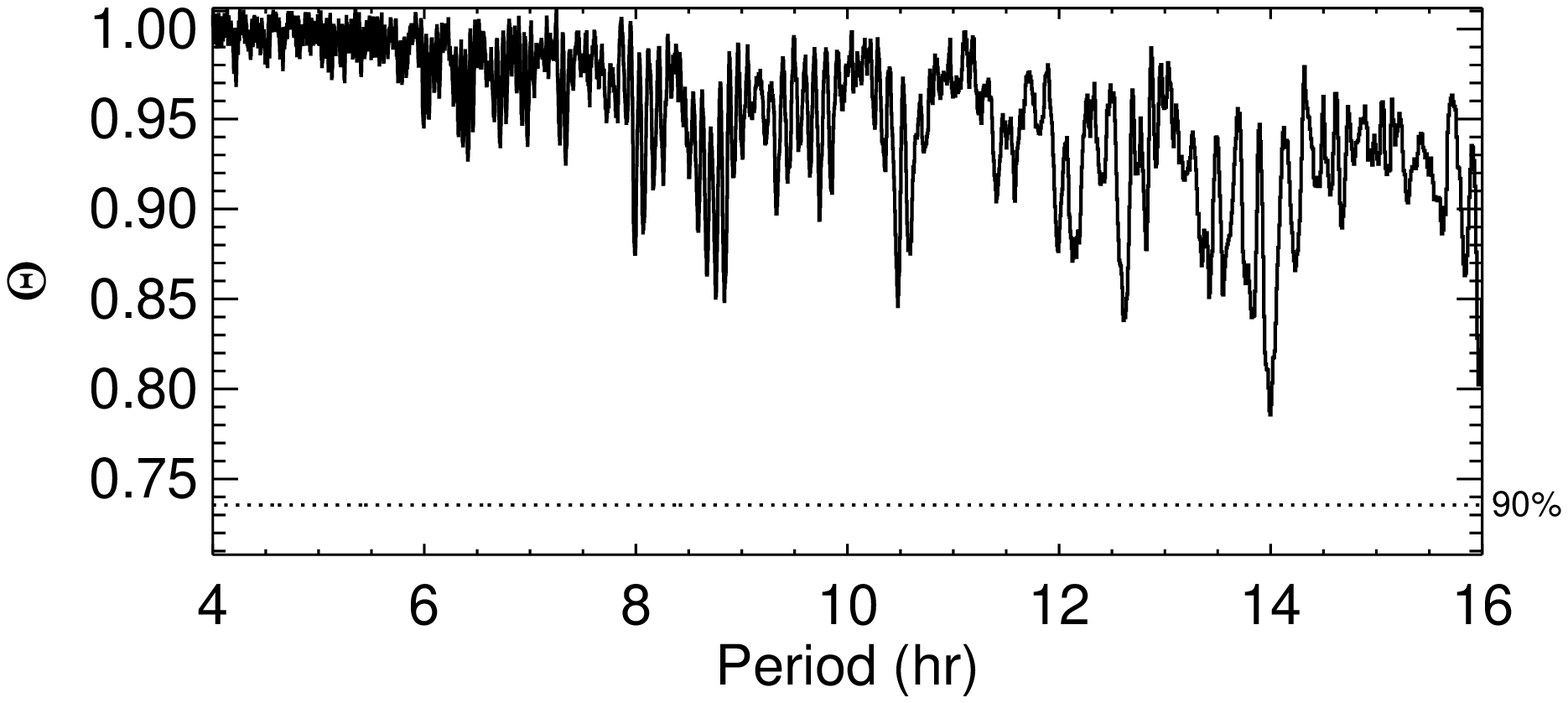} \\
\includegraphics[width=0.45\textwidth]{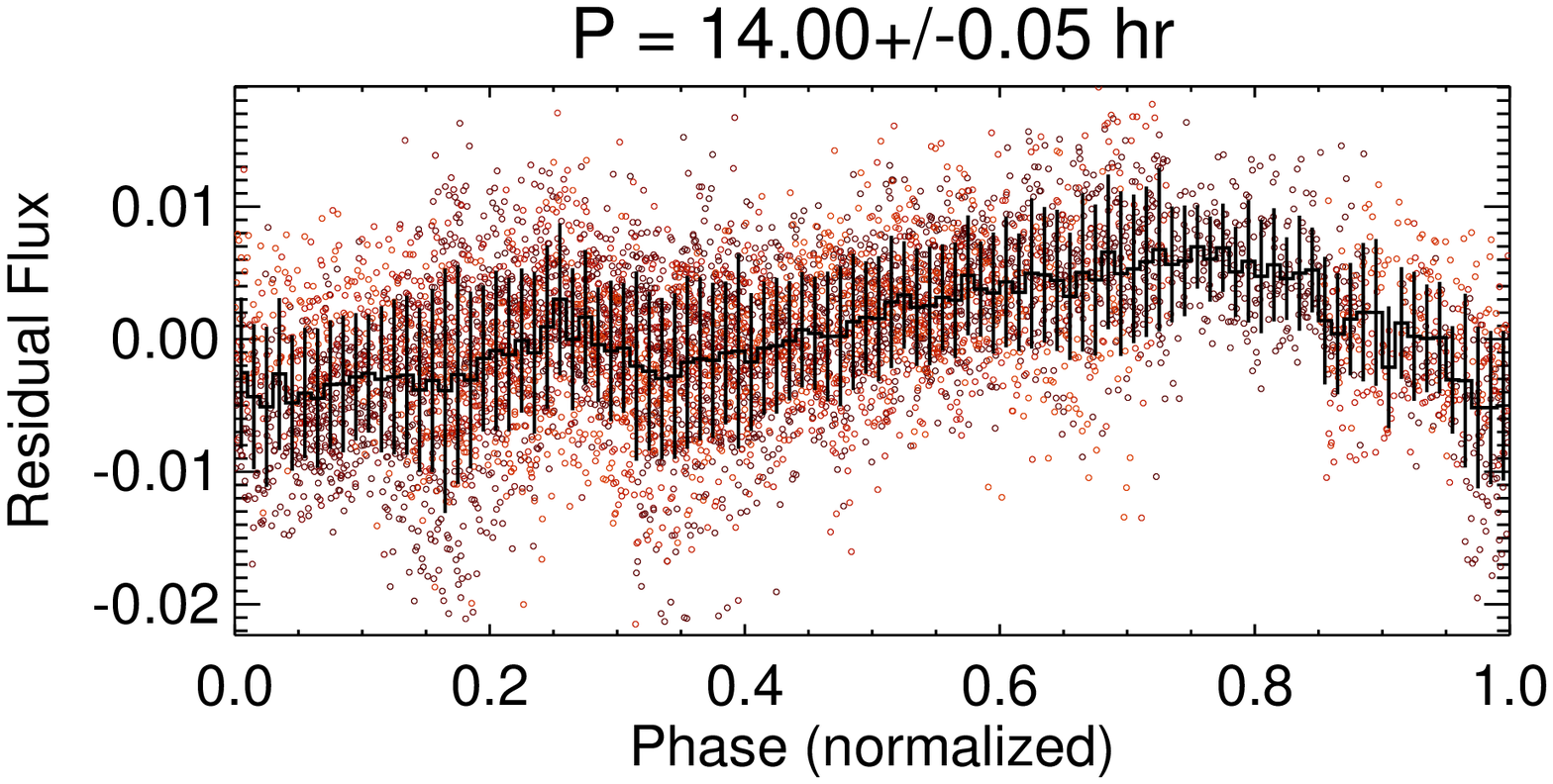} \\
\caption{(Top): Phased Dispersion Minimization statistic $\Theta$ as a function of period for the flare-cleaned, airmass-corrected, and daily-normalized TRAPPIST lightcurve of {\namesh}. A significance threshold of 90\% based on \citet{1997ApJ...489..941S} is indicated.  None of the features exceed this threshold.
(Bottom): Phased light curve for the strongest candidate period at 14.00$\pm$0.05~hr.
Individual measurements are indicated by small circles, where different shades of red indicate different cycles (17 in total). The black histogram with error bars (scatter per phase bin) delineates the mean phased light curve.
}
\label{fig:pdm}
\end{figure}

\begin{figure}[h]
\center
\epsscale{0.8}
\includegraphics[width=0.45\textwidth]{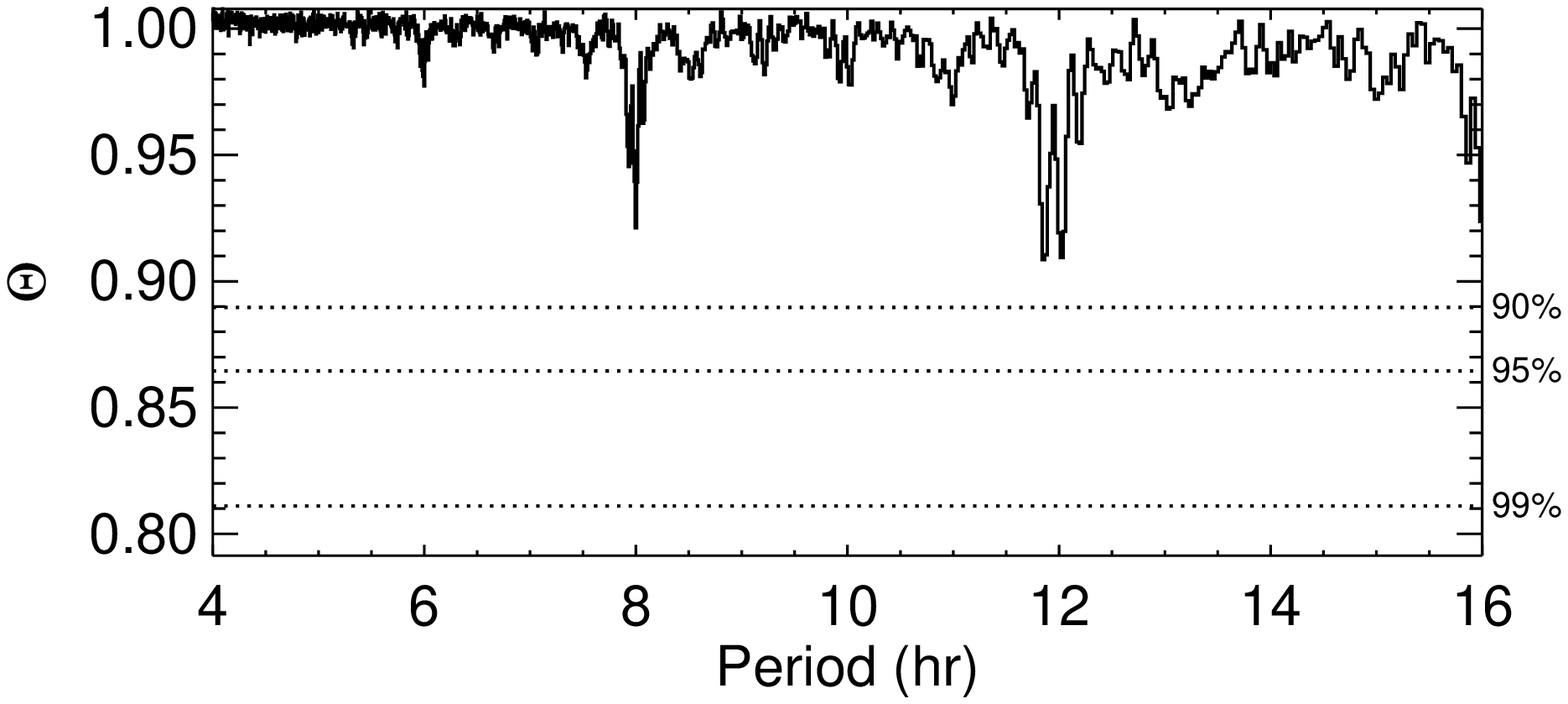} \\
\includegraphics[width=0.45\textwidth]{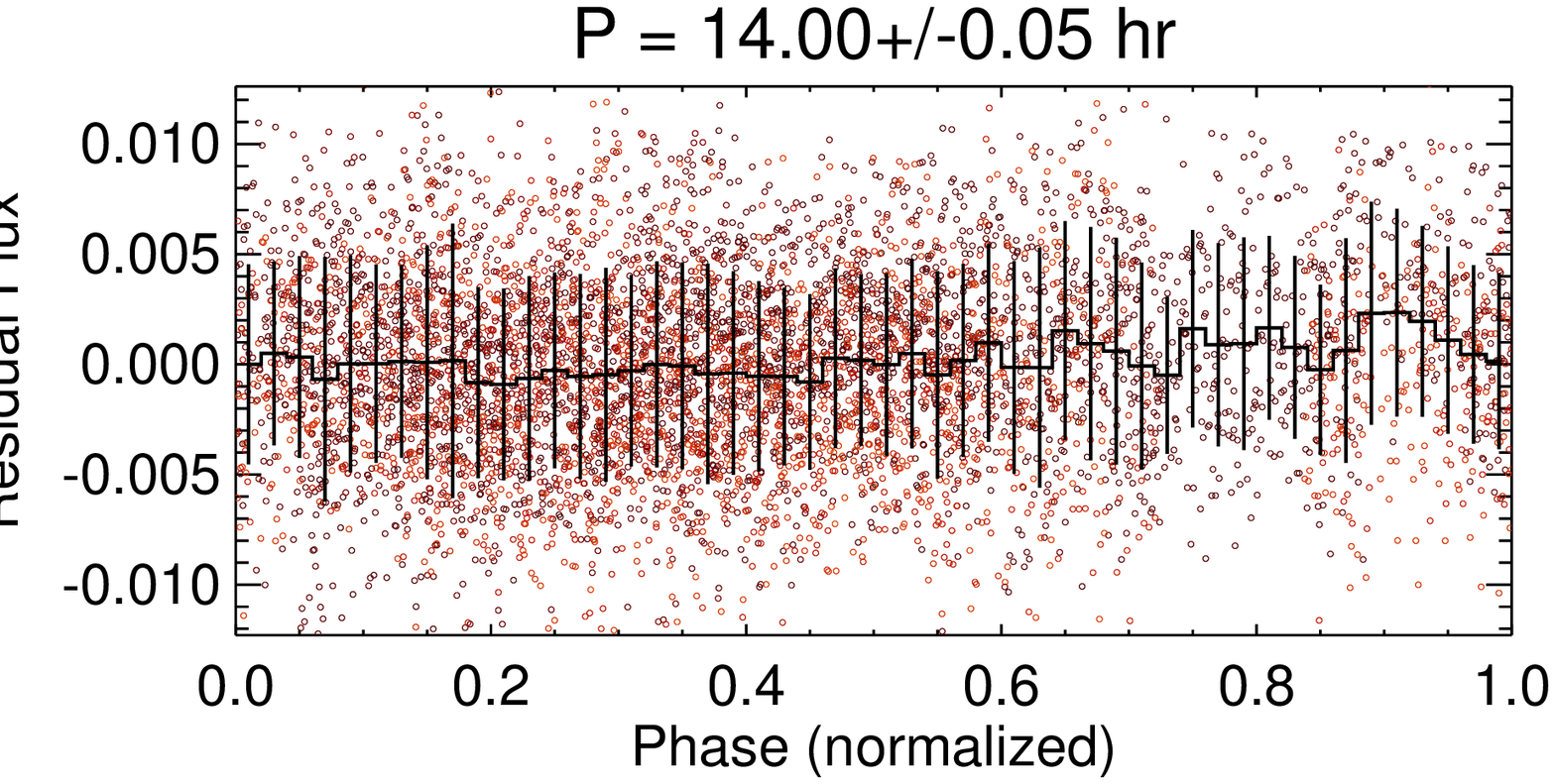} \\
\caption{Same as Figure~\ref{fig:pdm} but for the nearby comparison star 2MASS J07200688$-$0846504. The minima in $\Theta$ reflect the sampling window of the measurement (12~hr and harmonics) and are not significant.}
\label{fig:pdm-comp}
\end{figure}

The corrected and nightly-normalized light curve was analyzed using the Phase Dispersion Minimization technique \citep[PDM]{1978ApJ...224..953S,1990MNRAS.244...93D}.  Our implementation of this method is described in Appendix~A.  We examined periods between 4~hr and 16~hr, the lower limit consistent with the most rapidly rotating brown dwarfs and the upper limit chosen to be just above our rotational velocity limit. 
Figures~\ref{fig:pdm} and~\ref{fig:pdm-comp} display distributions of the PDM statistic $\Theta$ as a function of period for {\namesh} and the comparison star, respectively.
The PDM of {\namesh} shows considerable structure, but none of the features exceed our significance threshold of 90\%. The phased light curve of the strongest period at 14.00$\pm$0.05~hr displays a compelling, double-peaked pattern with an amplitude of 1.3$\pm$0.5\% (2.6$\sigma$). This period is consistent with a rotation axis inclination angle of 71$\degr$$\pm$8$\degr$. However, the $\chi^2$ of the phased photometric residuals relative to the smoothed lightcurve is formally consistent with a lightcurve without phasing. Several other
statistics were investigated, including Lomb-Scargle analysis \citep{1976Ap&SS..39..447L,1982ApJ...263..835S}, and trial sinusoidal fits to the data were also insufficient for identifying a robust period.  We therefore set a 3$\sigma$ upper limit of 1.5\% on periodic variability in this source, although larger episodic variations are clearly present.
Note that the comparison source also exhibits no significant period.

\section{The Binary Nature of {\namesh}}

\subsection{Identification as a Spectral Binary}

As described above, the SpeX spectrum of {\namesh} is best matched to that of the M9 spectral standard LHS~2924.
However, close inspection reveals specific discrepancies 
near 1.3~$\micron$ and 1.6~$\micron$, and a subtle ``dip'' feature at 1.62~$\micron$.
Such features have been previously noted in the combined-light spectra of 
very low mass spectral binary systems with late-type M or L dwarf primaries and T dwarf secondaries
(e.g., \citealt{2004ApJ...604L..61C,2007AJ....134.1330B,2010AJ....140..110G,2010ApJ...710.1142B,2011ApJ...739...49B,2012ApJ...757..110B,2012ApJ...753..156K,2013MNRAS.430.1171D,2014ApJ...792..119D,2014arXiv1408.3089B}). To assess whether {\namesh} is such a system, we used the fitting method described in \citet{2010ApJ...710.1142B}, comparing the spectrum of {\namesh} to 699 single star templates and 107,646 binary templates from the SPL. The latter were constructed using spectra of  M7-L4 dwarfs for the primary and L9-T7 dwarfs for the secondary, with component spectra scaled to absolute magnitudes using the $M_K$/spectral type relation of \citet{2007AJ....134.1162L}.  

Figure~\ref{fig:spex} displays the best fitting binary template, which consists of LHS~2924 paired to the T5 2MASS~J04070885+1514565 \citep{2004AJ....127.2856B}.
This combination reproduces the excess flux at 1.3~$\micron$, 1.6~$\micron$ and 2.1~$\micron$, as well as the shape of the 1.62~$\micron$ dip feature, and is a statistically significant better match to the data based on an F-test comparison ($>$99\% confidence).  Marginalizing over all fits weighted by the F-distribution, we infer component types of M9$\pm$0.5 and T5$\pm$0.7, and $\Delta{H}$ = 4.1$\pm$0.4, implying that the secondary contributes only 2\% of the combined light in this band.  This system is similar in composition to the previously confirmed M8.5 + T5 spectral binaries 2MASS~J03202839$-$0446358 (hereafter 2MASS~J0320$-$0446; \citealt{2008ApJ...681..579B,2008ApJ...678L.125B}) and SDSS~J000649.16$-$085246.3 (hereafter SDSS~J0006$-$0852; \citealt{2012ApJ...757..110B}).  

\subsection{Possible Detection of a Resolved Companion}

We searched for this potential companion using our NIRC2 $H$-band images.  Given the somewhat poor observing conditions, the wings of the PSF reach 2\% relative brightness at a relatively large radius of 0$\farcs$6 (4.2~AU; Figure~\ref{fig:image}). Only one point source is seen at larger radii, 4$\farcs$85 to the southwest of {\namesh} with $\Delta{H} \approx 6$. However, a stationary optical and near-infrared counterpart to this source is seen in both Digital Sky Survey and 2MASS images, the former going back to 1955, so it is likely an unrelated background star.  

To probe tighter separations, we self-subtracted the combined image after rotating by 180$\degr$ and offsetting to minimize the total squared deviation, effectively using the source PSF as its own model.  Figure~\ref{fig:image} shows that a faint source emerges upon this subtraction, 139$\pm$14~mas from the PSF center at a position angle of 262$\degr\pm$2$\degr$, corresponding to a projected separation of 0.84$\pm$0.17~AU at the 6.0$\pm$1.0~pc distance of this system.  The peak flux of this source relative to that of {\namesh} corresponds to a relative magnitude of 4.1$\pm$0.5~mag, consistent with that predicted from the SpeX spectral analysis, albeit with considerable uncertainty.  This source is therefore a promising candidate for the brown dwarf companion, although both its validity and physical association with {\namesh} must be confirmed.

\subsection{Limits on the Companion Orbit from Radial Velocity Monitoring}

We can constrain the presence of an unresolved companion from the radial velocity measurements obtained over 3 months with Keck/NIRSPEC and over 2 months with Lick/Hamilton Spectrograph.  In both sets of data, we find the measured radial velocities are constant within the uncertainties. For the NIRSPEC data, we measure a $\chi^2$ = 5.01 for 3 degrees of freedom, corresponding to a 17\% false alarm probability (marginal signficance); for the Hamilton data, we measure a $\chi^2$ = 5.21 for 6 degrees of freedom, corresponding to a 52\% false alarm probability (no significance).
The marginal significance in the NIRSPEC data may be indicative of an additional $<$0.5~{\kms} systematic error not included in our forward modeling analysis (see \citealt{2008ApJ...678L.125B,2012ApJ...757..110B}).

\begin{figure*}
\epsscale{1.1}
\includegraphics[width=0.32\textwidth]{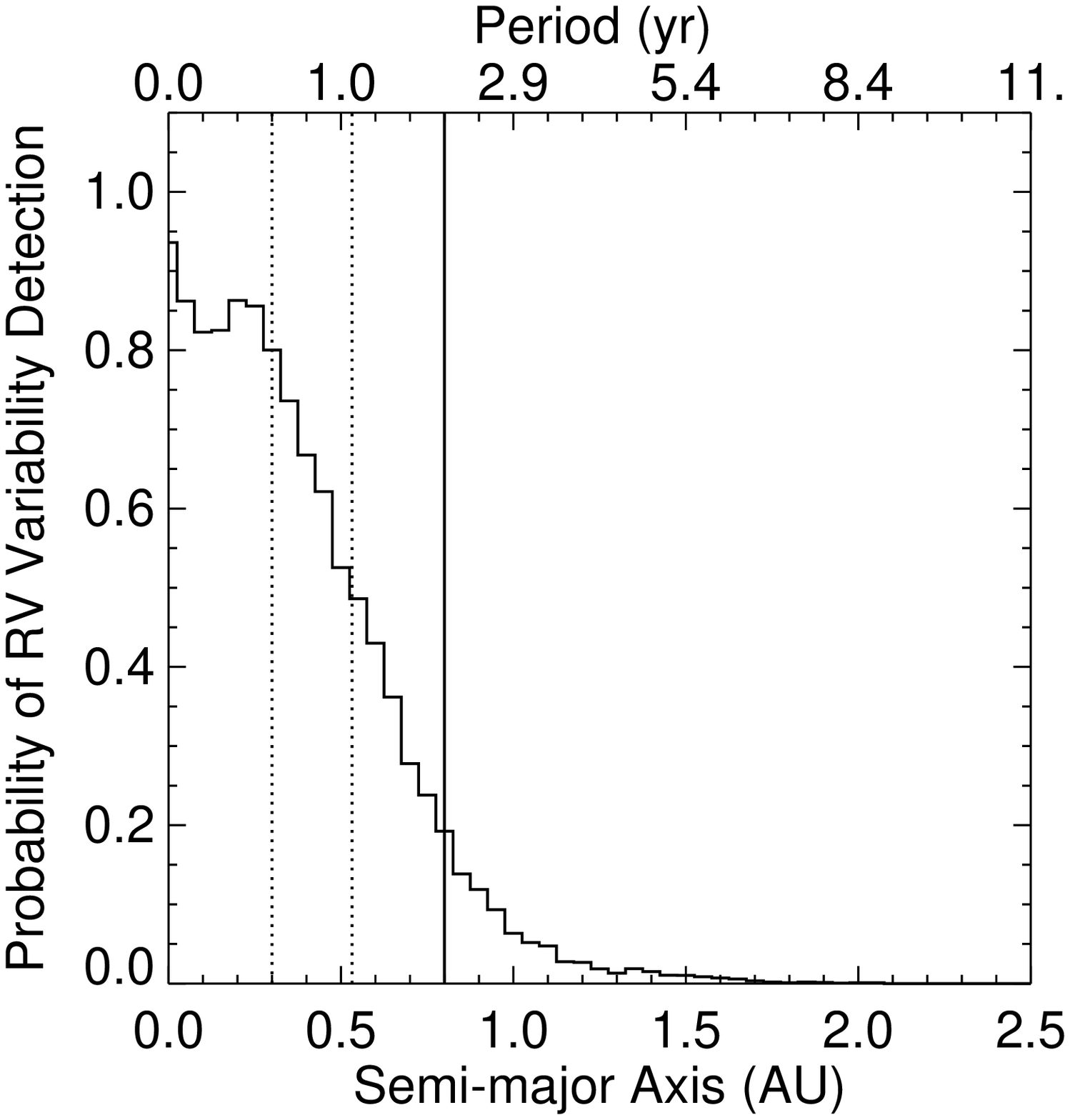}
\includegraphics[width=0.32\textwidth]{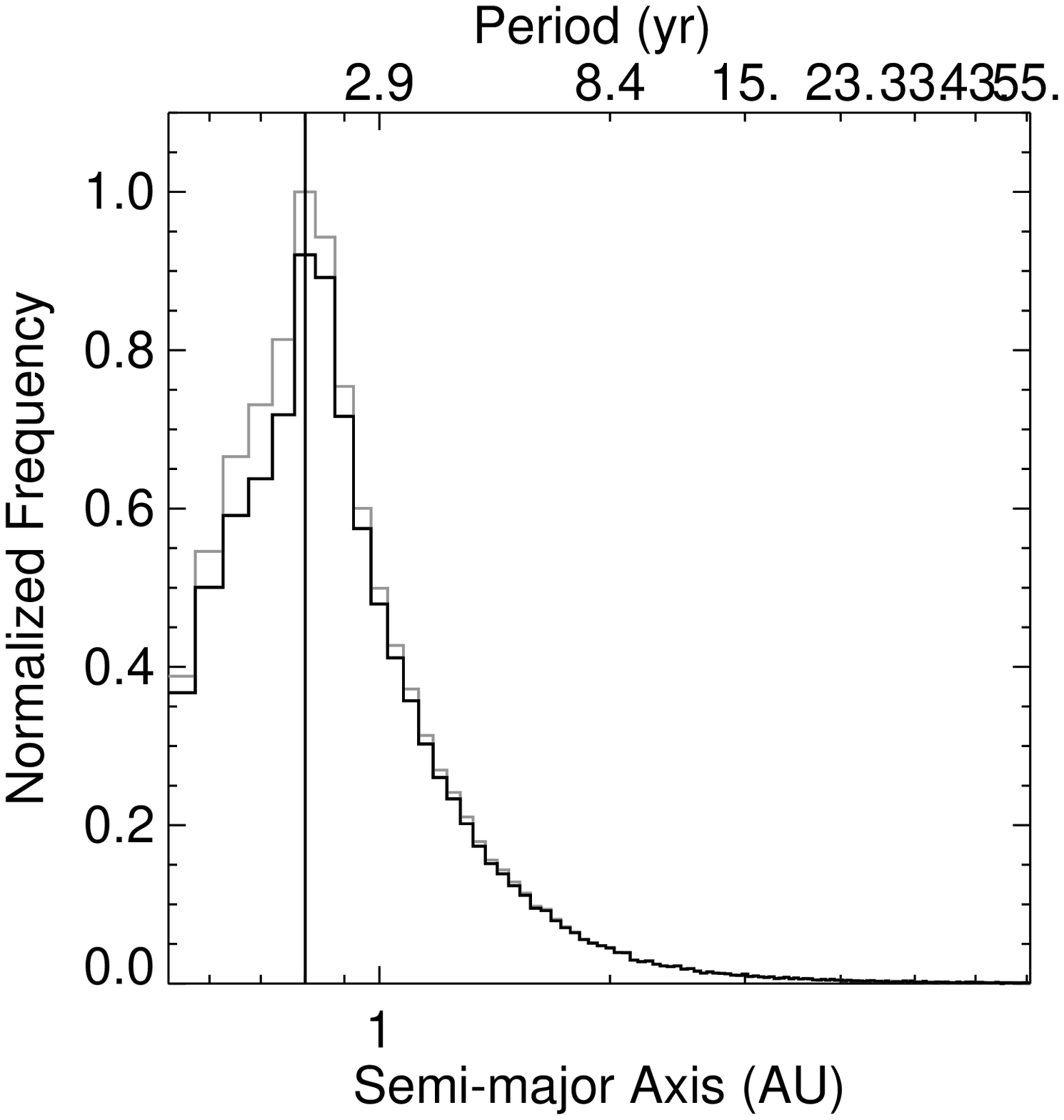} 
\includegraphics[width=0.32\textwidth]{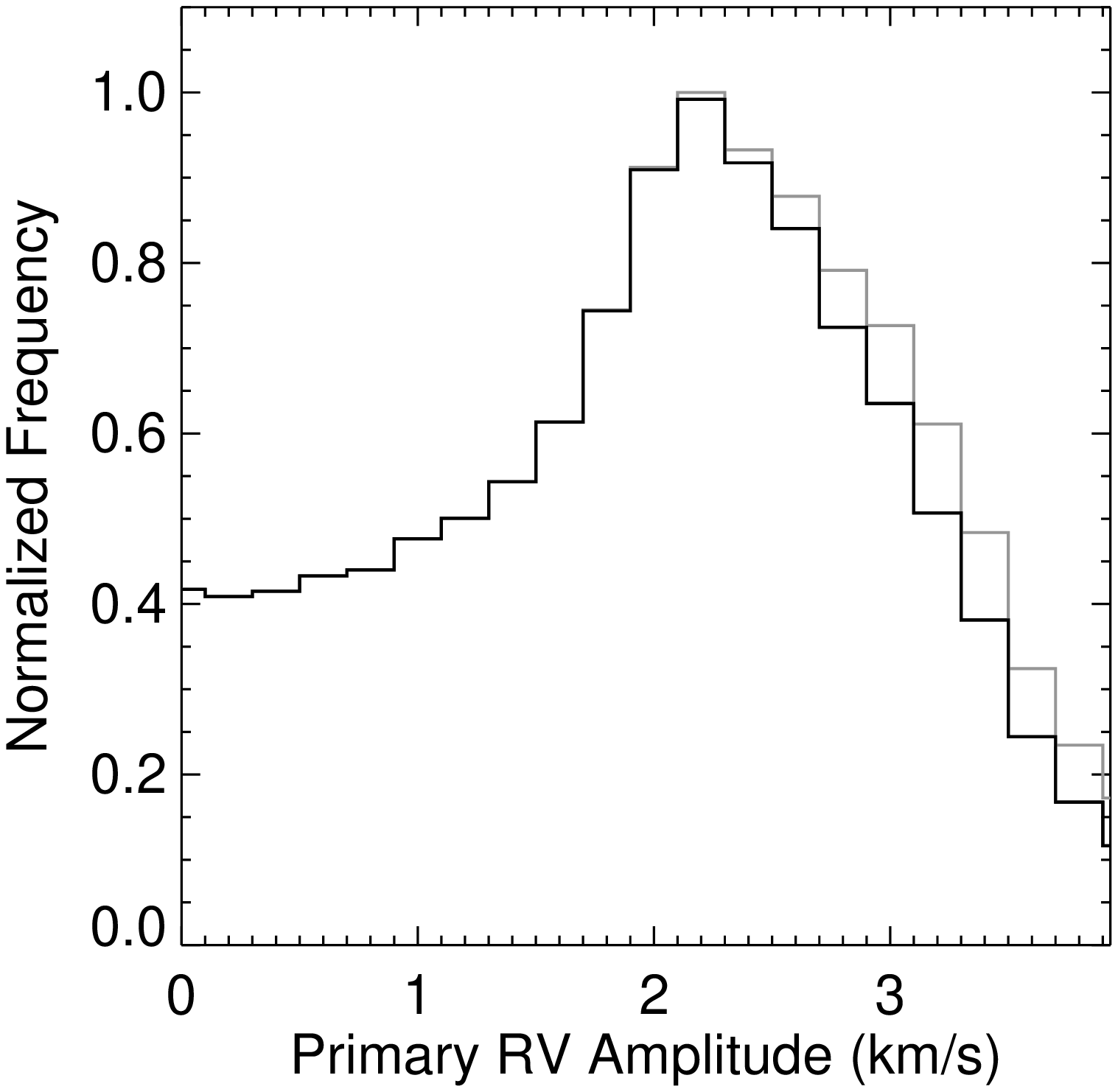} \\
\caption{Results of orbit simulations for the {\namesh}AB system based on the NIRSPEC radial velocity measurements and possible detection in NIRC2 images, assuming a system age of 1~Gyr.
{\em (Left)} Probability distribution of orbital semi-major axes based on the lack of variability in our radial velocity measurements.  The dotted lines indicate the 80\% and 50\% detection thresholds; the solid line indicates the most likely semi-major axis based on the putative NIRC2 detection (0.8~AU).
{\em (Center and Right)} Probability distributions of semi-major axis (center) and primary radial velocity variability amplitude (right)  based on the NIRC2 candidate detection alone (black histogram) and with the radial velocity constraints (grey histogram). 
\label{fig:rvsim}}
\end{figure*}

Using the simulations described in Appendix~B, we ascertained the range of semi-major axes ($a$)  ruled out assuming no detection of radial velocity variability. Given the slight offset between the NIRSPEC and Hamilton measurements, we treated the datasets separately.  We first converted the component spectral types to {\teff}s using the empirical relations of \citet{2009ApJ...702..154S}, and then converted these to masses using the evolutionary models of \citet{2001RvMP...73..719B} for ages 0.5-10~Gyr (Table~\ref{tab:rvsim}).
The NIRSPEC measurements rule out the presence of a companion with $a <$ 0.46--0.76~AU (0.26--0.44~AU) at the 50\% (80\%) confidence level, the range reflecting young to old ages (Figure~\ref{fig:rvsim}, Table~\ref{tab:rvsim}). The Hamilton data, given its shorter time coverage, provides a less stringent constraint.  Note that even the closest separations cannot be completely ruled out given the possibilities of a face-on orbit, an eccentric orbit observed at apoapse, or poor synching of observations and orbital inflection points.

\begin{deluxetable*}{ccccccccc}
\tablecaption{Constraints on the Companion Orbit from Radial Velocity and Imaging Measurements\label{tab:rvsim}}
\tabletypesize{\scriptsize}
\tablewidth{0pt}
\tablehead{
 & & & & \multicolumn{2}{c}{NIRSPEC only} & \multicolumn{3}{c}{NIRSPEC + NIRC2} \\
 \cline{5-6} \cline{7-9}
\colhead{System} &
\colhead{Primary\tablenotemark{a}} &
\colhead{Secondary\tablenotemark{a}} &
\colhead{Mass} &
\colhead{Detection} &
\colhead{Detection} &
\colhead{Semimajor} &
\colhead{Period} &
\colhead{Primary} \\
\colhead{Age} &
\colhead{Mass} &
\colhead{Mass} &
\colhead{Ratio} &
\colhead{Limit (80\%)} &
\colhead{Limit (50\%)} &
\colhead{Axis Range} &
\colhead{Range} &
\colhead{RV Range} \\
\colhead{(Gyr)} &
\colhead{({\msun})} &
\colhead{({\msun})} & & 
\colhead{(AU)} &
\colhead{(AU)} &
\colhead{(AU)} &
\colhead{(yr)} &
\colhead{({\kms})} \\
}
\startdata
0.5 & 0.066 & 0.027 & 0.41 & 0.26 & 0.46 &  0.77--1.2  & 2.2--4.3 & 1.2--2.4 \\
1.0 & 0.077 & 0.036 & 0.47 & 0.30 & 0.53 & 0.78--1.3  & 2.0--3.9 & 1.4--2.9 \\
5.0 & 0.084 & 0.065 & 0.77 & 0.41 & 0.72 & 0.77--1.2  & 1.8--3.4 & 2.2--4.6 \\
10.0 & 0.084 & 0.073 & 0.87 & 0.44 & 0.76 & 0.86--1.4 & 1.7--3.3 & 2.4--5.0 \\
\enddata
\tablenotetext{a}{Based on the evolutionary models of \citet{2001RvMP...73..719B}, system ages listed, and component {\teff}s of 2300~K and 1100~K for {\namesh}A and B  based on the {\teff}/spectral type relation of \citet{2009ApJ...702..154S}.}
\end{deluxetable*}

These limits are consistent with the location of the possible resolved companion.  We performed a second set of simulations that used both the radial velocity measurements and the observed separation to assess the distributions of probable semi-major axes, periods and primary radial velocity variability amplitudes. These are constrained to be in the range 0.77--1.4~AU, 1.7--4.3~yr and 1.2--5.0~{\kms}, respectively.  Both period and primary radial velocity amplitude are sensitive to the age and component masses of the system. Given the reasonably short time scale, maximum angular separation (0$\farcs$13--0$\farcs$23), and significant radial velocity perturbations, this system is an excellent target for individual component mass measurements in the near term.

\section{Discussion}

\subsection{Physical Properties of the {\namesh} System}

The inferred properties of {\namesh} are summarized in Table~\ref{tab:properties}.  Based on the analysis described above, we conclude that {\namesh}A is a relatively old, magnetically active, low-mass field star. Its degree of activity (e.g., flaring rate) and its rotation frequency are both somewhat below those of other late-type M dwarfs, suggesting some long-term angular momentum loss may have occurred for this source.  The putative T dwarf companion {\namesh}B must be substellar, and its time-dependent cooling permits a model-dependent constraint on the age of the system if the component masses can be determined. We can already estimate that the mass ratio of this system $q \equiv$ M$_2$/M$_1$ is highly age-dependent, with 0.47 $\leq q \leq$ 0.87 for 1 $\leq \tau \leq$ 10~Gyr. 
If the possible source detected in our NIRC2 images is the companion,
an inertial orbit measurement should be achieveable in 2--5~yr by combining astrometric and radial velocity monitoring.  Note that a robust detection of photometric variability would allow us to measure the rotational axis inclination of {\namesh}A and assess spin-orbit alignment in this system, a critical test of binary formation that has only been examined in one VLM binary to date \citep{2013A&A...554A.113H}.

It is worth noting the striking similarities between {\namesh}AB and two other late-M plus T spectral binaries SDSS~J0006$-$0852AB and 2MASS~J0320$-$0446AB, both M8.5 + T5 systems.  The common classification of the secondary in all of these systems is likely a selection bias, as this subtype lies at the peak of the so-called ``$J$-band bump'', the 1~$\micron$ brightening from late-L to mid-T likely caused by the depletion of photospheric clouds at the L dwarf/T dwarf transition \citep{2001ApJ...556..872A,2002ApJ...571L.151B,2013ApJ...772..129B,2008ApJ...685.1183L,2010ApJ...723L.117M}.
T5 companions are simply more readily detectable in blended light spectra due to their brighter magnitudes.  All three systems also have separations $\lesssim$1~AU, below the $\sim$4--7~AU peak of the separation distribution of resolved VLM binaries \citep{2007ApJ...668..492A}.  This agreement supports evidence that the spectral binary method is uncovering a significant population of tight binaries \citep{2014arXiv1408.3089B}.  Finally, all three systems appear to be relatively mature. SDSS~J0006$-$0852AB is $\gtrsim$7~Gyr based on the inactivity of its widely-separated M7 tertiary; 2MASS~J0320$-$0446AB is $\gtrsim$2~Gyr based on the mass and evolutionary state of its companion; {\namesh}AB may be $\sim$5~Gyr based on its old disk kinematics. This congruence may reflect the known preference for resolved VLM binaries to have nearly-equal mass components (\citealt{2003AJ....126.1526B,2003ApJ...586..512B,2003ApJ...587..407C}; however, see below). A brown dwarf closer to the hydrogen burning minimum mass (HBMM) must cool longer to reach a {\teff} $\approx$ 1200~K consistent with a mid-T dwarf.  From these few examples, we speculate that late-M dwarf plus T dwarf spectral binary systems specifically probe an old, tightly-bound population of VLM binaries, which are particularly useful systems for orbital mass measurements. 

\subsection{Late-M + T Dwarf Binaries in the Local Sample}

{\namesh} joins the M8.5 SCR 1845$-$6357AB system \citep{2006ApJ...641L.141B,2007A&A...471..655K} as one of two late-M plus T dwarf binaries in the immediate vicinity of the Sun.  Remarkably, these are the {\em only} binary systems among the 13 M7-M9.5 dwarf primaries known within 10 pc (Table~\ref{tab:latem}).  This fact is particularly surprising given the numerous efforts to identify faint companions to cool stars close to the Sun \citep{2003ApJ...587..407C,2003AJ....126.1526B,2006ApJ...644L..75M,2012AJ....144...64D}, and the apparent preference for high mass-ratio systems among VLM binaries.  As it stands, this volume-limited sample has a brown dwarf companion fraction, $\epsilon_{BD}$ = 15$^{+15}_{-5}$\%, that is marginally higher (but consistent with) its stellar companion fraction limit $\epsilon_*$ $<$12\% (1$\sigma$ binomial uncertainties). 

\begin{deluxetable*}{llcccl}
\tablecaption{M7--M9.5 Dwarf Primaries within 10~pc \label{tab:latem}}
\tabletypesize{\small}
\tablewidth{0pt}
\tablehead{
\colhead{Source} &
\colhead{SpT} &
\colhead{$\pi$} &
\colhead{$M_K$} &
\colhead{$V-K$} &
\colhead{Ref.} \\
 & & 
\colhead{(mas)} \\
}
\startdata
SCR~1845-6357AB & M8.5+T6 & 259.5$\pm$1.1 & 10.58 & 8.89 & 1,10 \\
DENIS~J1048-3956 & M8.5 & 249.8$\pm$1.8  & 10.44  & 8.88  & 1  \\
LSR~J1835+3259 & M8.5 & 176.5$\pm$0.5  & 10.41  & 9.10 & 2 \\
{\bf {\namesh}AB} & M9.5+T5 & 166$\pm$28  & 10.5  & \nodata & 3, 4 \\
LP~944-20 & M9 & 155.9$\pm$1.0  & 10.51 & 9.15 & 5 \\
GJ~3877  & M7 & 152$\pm$2  & 9.84 & 8.02  &  5, 7 \\
SCR~1546-5534 & M8: & 149$\pm$40 & 10.0 &\nodata  & 6 \\
SIPS~1259-4336 & M8: & $\sim$128 &  $\sim$10.1  & \nodata  & 9, 11  \\
LEHPM~1-3396 & M9 & 121$\pm$4 & 10.80  & \nodata & 12, 13 \\
LHS~2065 & M9 & 118.0$\pm$0.8  &  10.30 & 9.00 & 5, 7 \\
1RXS~J1159-5247 & M9 & 105.54$\pm$0.12  & 10.44  & \nodata & 14, 15 \\
LP~655-48 & M7 & 106$\pm$3 & 9.66  & 8.31  & 1, 16 \\
LP~647-13 & M9 & 104$\pm$2  & 10.52  & 8.84 & 8  \\
\enddata
\tablerefs{(1) \citet{2004AJ....128.2460H}; (2) \citet{2003AJ....125..354R}; 
(3) \citet{2014A&A...561A.113S}; (4) This paper; 
(5) \citet{2014AJ....147...94D}; (6) \citet{2011AJ....142...92B}; (7) \citet{1995AJ....109..797K}; (8) \citet{2005AJ....130..337C}; (9) \citet{2005A&A...435..363D}; (10) \citet{2005AJ....129..409D}; (11) T.~Henry, priv.~comm.; (12) \citet{2006MNRAS.366L..40P}; (13) \citet{2012ApJ...752...56F}; (14)  \citet{2004A&A...415..265H}; (15) \citet{2014A&A...565A..20S}; (16) \citet{2012ApJ...758...56S}.}
\end{deluxetable*}

\begin{figure}[h]
\center
\epsscale{1.1}
\plottwo{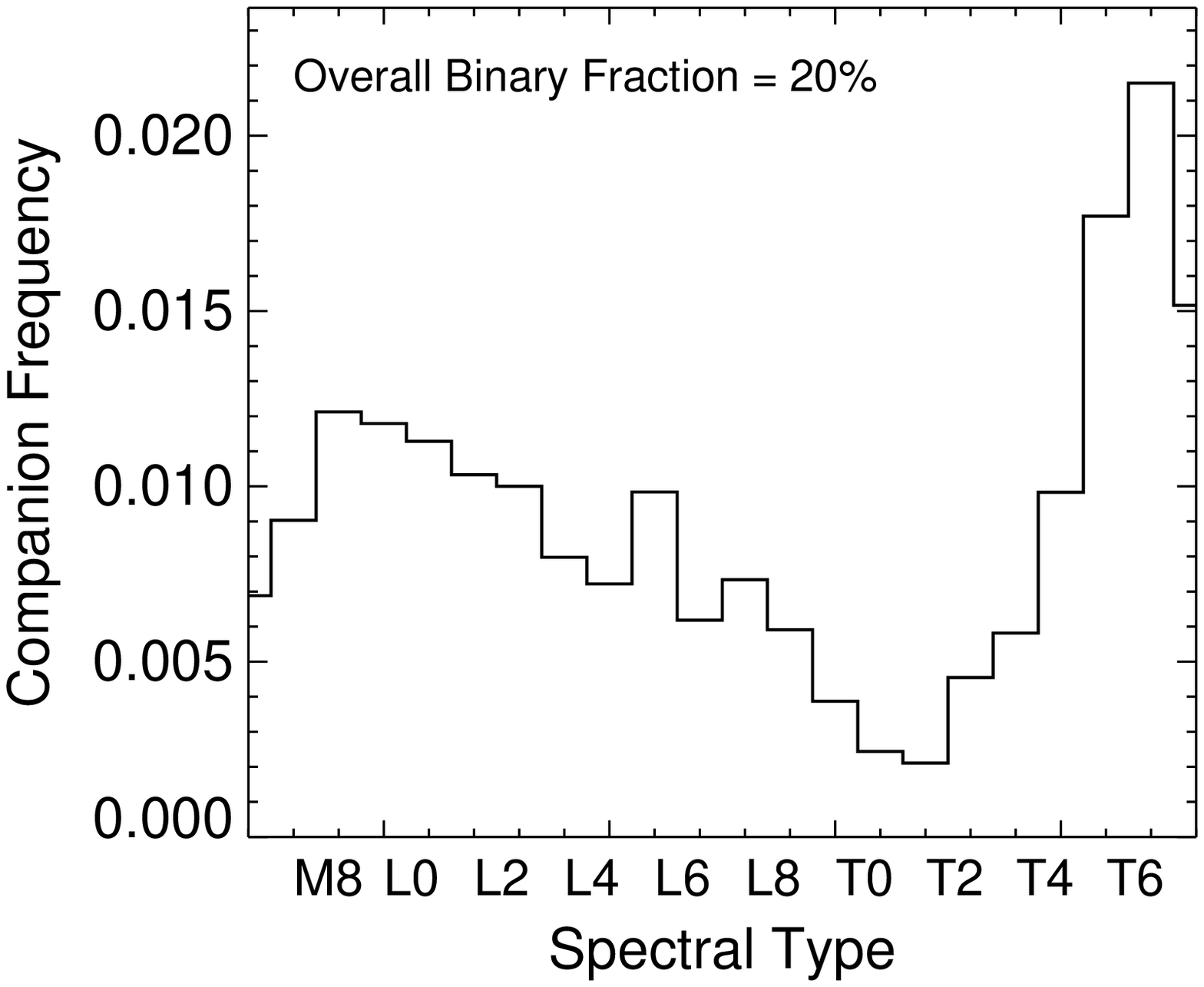}{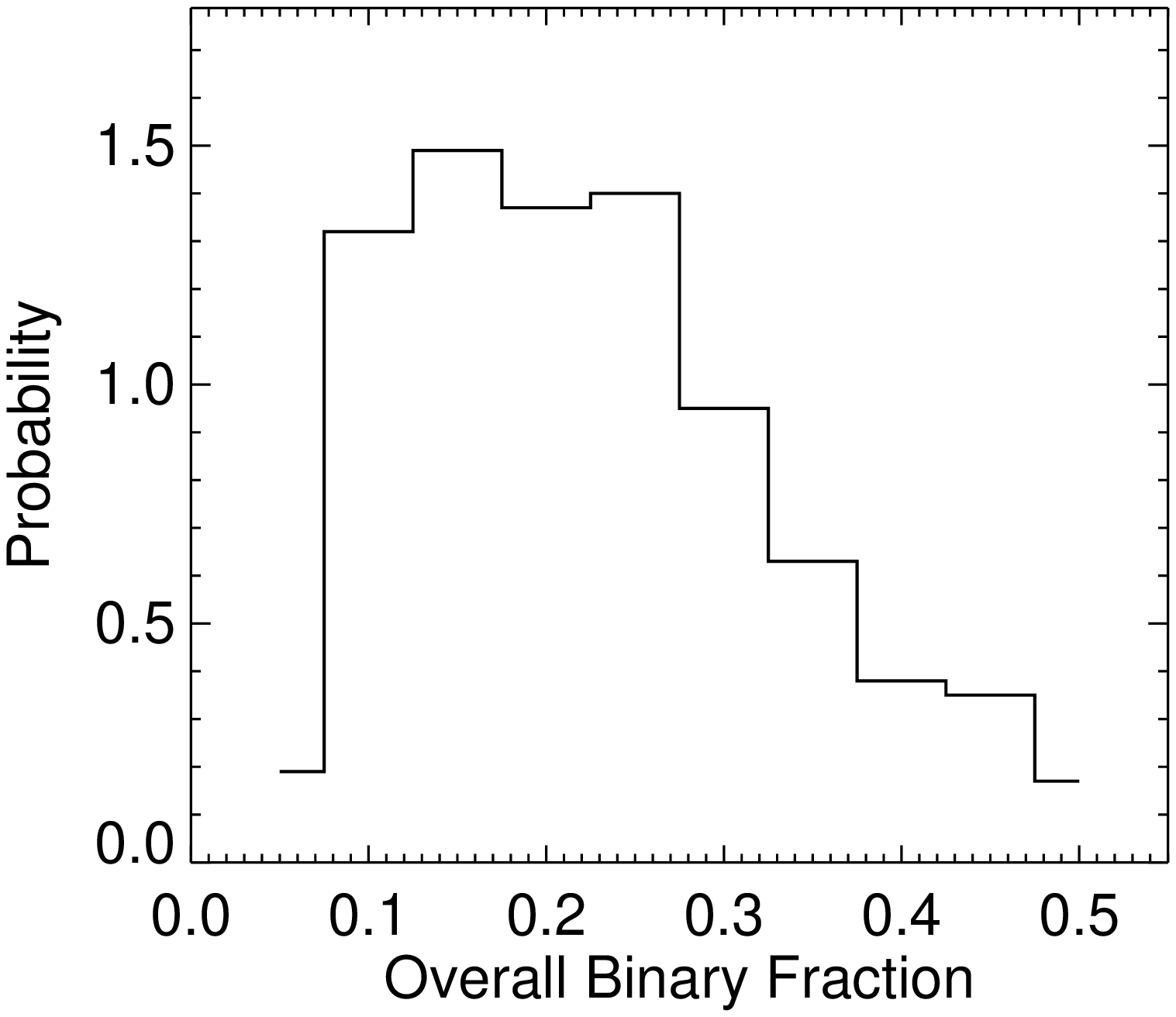}
\caption{(Left): Frequency distribution of companion spectral types for M7--M9.5 primaries based on the simulation described in Appendix~C and assuming an overall binary fraction $\epsilon$ = 20\%. Mid-type T dwarfs are far more common per subtype than late M and L dwarfs, and comprise $\sim$25\% of all companions.  
(Right): Probability of finding two or more exclusively T2--T7 companions in a sample of 13 late M dwarfs as a function of overall binary fraction, based on these simulations.  The distribution plateaus around 1.4\% for 10\% $< \epsilon <$ 25\%. The decline at smaller $\epsilon$ reflects the lack of companions; the decline of large $\epsilon$ results from contamination from earlier-type companions.
}
\label{fig:binfrac}
\end{figure}

Is the nature of multiples in the local late-M dwarf population simply due to small number statistics? To examine this question, we performed a population simulation similar to those described in \citet{2004ApJS..155..191B,2007ApJ...659..655B}, combined with a random draw experiment to determine the likelihood of various binary configurations; see Appendix~C for details. From our simulations we find that T dwarfs are remarkably common companions to late M dwarfs, comprising $\sim$25\% of secondaries (Figure~\ref{fig:binfrac}).  Their relative abundance derives from two factors.
First, the typical masses of M7-M9.5 dwarfs in our simulation, 0.089$\pm$0.003~{\msun}, are close to the HBMM, so stellar companions exist over a relatively narrow range of masses. Brown dwarf companions to late-M primaries consistently outnumber stellar companions in our simulation; in contrast, only $\sim$9\% of systems with mid-M (M4--M7) primaries host T dwarf companions.  Second, there is the well-known ``pile-up'' of T dwarfs in field brown dwarf populations due to their slow cooling rates \citep{2004ApJS..155..191B,2005ApJ...625..385A}. T dwarfs are the most common companion to the oldest systems in our simulation ($>$5~Gyr). This is consistent with the old ages inferred for SDSS~J0006$-$0852AB, 2MASS~J0320$-$0446AB and {\namesh}AB.

Even with the prevalence of T dwarf companions to late-M primaries, the incidence of finding 2 or more T dwarf companions in the nearby sample--and only T dwarf companions--remains low, plateauing around 1.4\% for a total binary fraction 10\% $<$ $\epsilon$ $<$ 25\%.  The local sample would seem to be an anomaly if the underlying multiplicity distributions are accurately characterized.  It is possible that more (and less) massive companions to other late M dwarfs in the nearby sample have yet to be found, and may be uncovered astrometrically with the current GAIA mission \citep{2001A&A...369..339P,2004RMxAC..21..251Z}. We also note that while this specific configuration is rare, there are roughly 10$^8$ 10-pc ``bubbles'' in the 3--12~kpc Galactic disk, making our sample of M dwarfs one of a million.

Finally, we note that the peak probability of exclusively T dwarf companions to late M dwarfs occurs in a range of $\epsilon$ consistent with previous determinations of the VLM binary fraction \citep{2005ApJ...621.1023S,2006ApJS..166..585B,2006AJ....132..663B,2007ApJ...668..492A}. This is significant, as the decline in frequency at high fractions is caused by increased contamination by earlier-type secondaries. The lack of such companions in the nearby sample may be a consequence of the low binary fraction of VLM dwarfs compared to more massive stars.  The companionship of late-M dwarfs appears to be determined by the confluence of four key statistics of VLM dwarfs: the mass function across the substellar limit, the cooling rate of brown dwarfs, the mass ratio distribution of VLM dwarfs, and the underlying multiplicity fraction.  The importance of these statistics for brown dwarf formation and interior theories motivates a complete assessment of the multiplicity properties of late M dwarfs in a much larger volume.

\section{Summary}

We have conducted a detailed investigation of the recently identified, nearby VLM dwarf {\namesh}.
From optical and infrared imaging and spectroscopic investigations, we have determined the optical and near-infrared classifications of this source, improved its astrometry, measured its spatial kinematics and rotational velocity, identified persistent and flaring magnetic activity, and found evidence for a brown dwarf companion. 
The reduced level of activity in this source compared to other late-M dwarfs, and its old disk kinematics, both suggest that this is a relatively mature system.  The putative T dwarf companion, identified in combined-light spectroscopy and possibly resolved at 1~AU projected separation, will aid in constraining the age, as astrometric and radial velocity monitoring over the next few years should allow us to map the orbit of the system and extract individual component masses.  Remarkably, this is one of only two binaries among late M dwarfs in the immediate vicinity of the Sun, and both have T dwarf companions.  We argue that, while rare, this may reflect a combination of the proximity of the primary to the HBMM, the evolutionary properties of brown dwarfs, and the underlying binary fraction, making multiplicity studies of late M dwarfs a potentially useful window into brown dwarf formation and evolution theories.   

\acknowledgements
The authors thank 
Bill Golisch and John Rayner at IRTF; 
Wayne Earthman, Erik Kovacs, Donnie Redel and Pavl Zachary at Lick Observatory;
Diane Harmer and Krissy Reetz at KPNO; 
Scott Dahm, Greg Doppmann, Heather Hershley, Gary Punawai, Luca Rizzi, and Terry Stickel at Keck for their assistance with the observations.
We also acknowledge useful discussions with Gregg Hallinan and Stuart P.\ Littlefair on
M dwarf magnetic activity; and 
John Gizis, Todd Henry, J.\ Davy Kirkpatrick and I.\ Neill Reid on the 10~pc sample.
We thank our referee, R.\ Scholz, for his very helpful comments that allowed us to considerably improve the manuscript.
This research has made use of the SIMBAD database,
operated at CDS, Strasbourg, France;
the M, L, T, and Y dwarf compendium housed at DwarfArchives.org;
and the SpeX Prism Spectral Libraries at \url{http://www.browndwarfs.org/spexprism}.
C.M. acknowledges support from the National Science Foundation under award No.\ AST-1313428. 
TRAPPIST is a project funded by the Belgian Fund for Scientific Research (F.R.S.-FNRS) under grant FRFC 2.5.594.09.F, with the participation of the Swiss National Science Foundation. 
M.\ Gillon and E.\ Jehin are F.R.S.-FNRS Research Associates. 
L.\ Delrez and J.\ Manfroid acknowledge the support of the F.R.S.-FNRS for their PhD theses.
The authors wish to recognize and acknowledge the very significant cultural role and reverence that the summit of Mauna Kea has always had within the indigenous Hawaiian community.  We are most fortunate to have the opportunity to conduct observations from this mountain.

\appendix

\section{Phase Dispersion Minimization Analysis}

The Phase Dispersion Minimization (PDM) technique identifies periodic signals in time-series data by searching for minimal dispersion about a mean phased signal \citep{1978ApJ...224..953S}.
The frequency sampling is set by the fundamental frequency $f_1$ $\equiv$ 1/$T$ = 0.021 cycles~day$^{-1}$, where $T$ = 1160~hr is the full monitoring period \citep{1982ApJ...263..835S,1988ComPh...2...77P}.
We examined 1081 periods in the period range 4~hr $< P <$ 16~hr, sampled evenly in frequency space in steps of $f_1/10$.
For each period, we phase-folded the lightcurve, computed a mean curve sampled at $r$ = 100 linearly-spaced phase points, then computed the $\chi^2$ deviation of the phased data relative to the mean curve.  The statistic of merit is the ratio
\begin{equation}
\Theta(P) = \frac{\chi^2(P)}{\chi^2_o}\frac{N-1}{N-r}
\end{equation}
where $\chi^2_o$ is measured from the unphased lightcurve, and $N$ = 5895 is the total number of data points.

Following \citet{1997ApJ...489..941S}, we assessed the significance of minima in $\Theta(P)$ using the regularized incomplete beta function $I$:
\begin{equation}
{\rm Pr}(\Theta < z) = \left[I_{\frac{N-r}{N-1}z}(\frac{N-r}{2},\frac{r-1}{2})/I_1(\frac{N-r}{2},\frac{r-1}{2})\right]^{1/N_f}
\end{equation}
Here, Pr$(\Theta < z)$ is the false-alarm probability when $\Theta$ is less than some value 0 $< z <$ 1 by chance; 1-Pr gives the significance of a period corresponding to a minimum in $\Theta$.  The 1/$N_f$ exponent is the band penalty incurred when searching for multiple periods, where we have used 
\begin{equation}
N_f = 2\frac{f_{high}-f_{low}}{f_{1}} = 432
\end{equation}
as the number of independent frequencies sampled \citep{1988ComPh...2...77P}. 
Periods with 1-Pr $>$ 90\% were deemed significant, but no features in the cleaned and normalized light curve satisfied this limit.

\section{Simulations for Orbit Constraints}

To quantify the companion detection limits from our radial velocity observations, and determine whether the candidate companion identified in NIRC2 imaging data is consistent with these limits, we performed a pair of Monte Carlo orbit simulations.  The first simulation aimed to determine the range of orbital semimajor axes that would have been detected given the radial velocity measurement uncertainties and sampling.  We generated a large number (10$^6$) of hypothetical orbits, uniformly sampling semi-major axes 0 $< a <$ 2.5~AU, eccentricities 0 $< \epsilon <$ 0.6 \citep{2011ApJ...733..122D}, inclinations 0 $< i <$ $\pi$, longitude of ascending node 0 $< \Omega <$ 2$\pi$, argument of periapse 0 $< \omega <$ 2$\pi$, and mean anomaly angle 0 $< M_{NIRC2} <$ 2$\pi$ for the NIRC2 imaging epoch.  We then solved for the maximum radial velocity amplitude of the primary component (e.g., \citealt{2005AJ....129.1706F}):
\begin{equation}
V_1 = \frac{2\pi{a}\sin{i}}{P\sqrt{1-\epsilon^2}}\frac{M_2}{M_1+M_2},
\end{equation}
with component masses M$_1$ and M$_2$ estimated from the evolutionary models of \citet{2001RvMP...73..719B}, system ages of 0.5, 1, 5 and 10~Gyr, and  component {\teff}s of 2300~K and 1100~K for {\namesh}A and B  based on the {\teff}/spectral type relation of \citet{2009ApJ...702..154S} (Table~\ref{tab:rvsim}).  The observed radial motion of the primary at each NIRSPEC or Hamilton epoch $t_i$ (these datasets were modeled separately) was then calculated as
\begin{equation}
v(t_i) = V_1(\epsilon\cos{\omega} + \cos{(T(t_i)-\omega)})
\end{equation}
where 
\begin{equation}
\tan{\frac{T(t_i)}{2}} = \sqrt{\frac{1+\epsilon}{1-\epsilon}}\tan{\frac{E(t_i)}{2}}
\end{equation}
relates the true anomaly $T$ to the eccentric anomaly $E$, which is in turn related to the mean anomaly through Kepler's Equation:
\begin{equation}
M(t_i) - M_0 = 2\pi\frac{t_i-\tau_0}{P} = E - \epsilon\sin{E}
\end{equation}
Here, $P$ is the period of the orbit, determined from the estimated masses, input semimajor axis, and Kepler's Period Law; and $t_i-\tau_0$ is the time since periastron passage.  Setting $M_0$ = 0 and $t_{NIRC2}$ = 0 (the time of the NIRC2 image), we can set $\tau_0 = -PM_{NIRC2}/2\pi$ and solve for the primary velocities numerically.  

For each simulated orbit, we calculated $\chi^2$ for the radial velocity epochs relative to their mean, using the corresponding observational uncertainties. We then determined the fraction of orbits that exceeded our measured $\chi^2$ as a function of semi-major axis.  Figure~\ref{fig:rvsim} displays the results of this calculation for the NIRSPEC data and a system age of 1~Gyr.  Due to inclination variations and sampling, detection probabilities never reach 100\%; we therefore use 50\% and 80\% probabilities of detection as our thresholds. 

To add in the constraint of a possible detection of the companion at a projected separation of $\rho = 0\farcs14$, we used the same orbital parameters to assess the range of semi-major axes consistent with this separation.  Cartesian positions in the plane of the sky for an orbit of unit semi-major axis are
\begin{align}
x &  =  A(\cos{E}-\epsilon) + F\sqrt{1-\epsilon^2}\sin{E} \\
y & =  B(\cos{E}-\epsilon) + G\sqrt{1-\epsilon^2}\sin{E} \\
\end{align}
were $A$, $B$, $F$ and $G$ are the Thiele-Innes constants \citep{1907Obs....30..310I,1927BAN.....3..261V}:
\begin{align}
A & = \cos{\omega}\cos{\Omega} - \sin{\omega}\sin{\Omega}\cos{i} \\
B & = \cos{\omega}\sin{\Omega} + \sin{\omega}\cos{\Omega}\cos{i} \\
F & = -\sin{\omega}\cos{\Omega} - \cos{\omega}\sin{\Omega}\cos{i} \\
G & = - \sin{\omega}\sin{\Omega} + \cos{\omega}\cos{\Omega}\cos{i} \\
\end{align}
The total projected separation $\Delta = \sqrt{x^2+y^2}$ can then be compared to the observed projected separation ($\rho$d) to constrain the semimajor axis for a given orbit $i$:
\begin{equation}
a_i = \frac{\rho{d}}{\Delta_i}
\end{equation}
Figure~\ref{fig:rvsim} displays the distributions of semi-major axes and primary radial velocity variability amplitudes consistent both with the possible detection and lack of detectable radial velocity variation.  

\section{Simulation for Local Binary Statistics}

We performed another Monte Carlo simulation to assess the likelihood of finding two or more T dwarf companions (exclusively) in the sample of 13 M7--M9.5 dwarfs within 10~pc of the Sun (Table~\ref{tab:latem}). We started with a sample of $N$ = 10$^5$ primaries with uniform ages spanning 0.5--7~Gyr (corresponding to the typical ages of disk stars) and masses 0.013~{\msun} $\leq$ M $\leq$ 0.2~{\msun} distributed as a power-law mass function $dN/dM \propto M^{-0.5}$ \citep{1999ApJ...521..613R,2012ApJ...753..156K,2013MNRAS.433..457B}.  For each system we assigned a secondary mass using the power-law mass ratio distribution of \citet{2007ApJ...668..492A}, $dN/dq \propto q^{1.8}$, based on Bayesian analysis of VLM imaging samples; this distribution favors equal-mass systems.  The component masses and system ages were transformed to bolometric luminosities using the evolutionary models of \citet{2001RvMP...73..719B}, and these converted into spectral types by combining the spectral type/absolute $J$-band magnitude relation of \citet{2012ApJS..201...19D} with the spectral type/$J$-band bolometric correction relation of \citet{2010ApJ...722..311L}.  
Systems with component spectral types outside the range M5-T9 were rejected. 

We selected a subset of M7-M9.5 sources (4626 systems) and used these as the primaries for our random-draw experiment. The distribution of companion types for this sample is shown in Figure~\ref{fig:binfrac}. Flagging a randomly-assigned subset $N_b = \epsilon{N}$ of these systems as actual binaries, with $\epsilon \in$ [0.05,0.5], we repeatedly (10$^5$ times) drew 13 systems from this collection and determined the fraction that contained at least two companions with spectral types T2--T7 (detectable by the spectral binary method; \citealt{2007AJ....134.1330B}) and no other companion types.
Figure~\ref{fig:binfrac} displays the frequency of this outcome as a function of overall binary fraction.


\end{document}